\documentclass[aps,prb,twocolumn,superscriptaddress,showpacs]{revtex4-1}
\usepackage{amsmath,amssymb,graphicx}
\usepackage{float}

\usepackage[T1]{fontenc}
\usepackage{chancery}

\usepackage{xcolor}
\usepackage{physics}
\usepackage[bottom]{footmisc}

\IfFileExists{newtxtext.sty}   {\usepackage{newtxtext,newtxmath}}
{\IfFileExists{stix.sty}
	{\usepackage{stix}}
	{\IfFileExists{mathptmx.sty}
		{\usepackage{mathptmx}}{} } }

\usepackage{textcomp}

\usepackage{bm}
\usepackage{mathbbol}

\IfFileExists{siunitx.sty}{\usepackage{booktabs,siunitx}}{}

\usepackage{color}
\definecolor{LinkColor}{rgb}{0.256,0.439,0.588}
\usepackage{hyperref}
\hypersetup{
	pdfauthor={Tsung-Cheng Lu},
	colorlinks=true,
	citecolor=LinkColor,
	linkcolor=LinkColor,
	urlcolor=LinkColor
}

\usepackage{pifont}
\usepackage{epstopdf}
\usepackage{caption}
\usepackage{subcaption}
\usepackage{simplewick}
\renewcommand{\raggedright}{\leftskip=0pt \rightskip=0pt plus 0cm}
\let\vec\mathbf

\newcommand{\be}{\begin{equation}}
\newcommand{\ee}{\end{equation}}

\newcommand{\beq}{\begin{eqnarray}}
\newcommand{\eeq}{\end{eqnarray}}

\begin{document}

    \title{Spacetime duality between localization transitions and measurement-induced transitions
}
\author{Tsung-Cheng Lu} 
\affiliation{Department of Physics, University of California at San
	Diego, La Jolla, California 92093, USA}
	\author{Tarun Grover}
\affiliation{Department of Physics, University of California at San
		Diego, La Jolla, California 92093, USA}

	\begin{abstract}
		\noindent
	Time evolution of quantum many-body systems typically leads to a state with maximal entanglement allowed by symmetries. Two distinct routes to impede entanglement growth are inducing localization via spatial disorder, or subjecting the system to non-unitary  evolution, e.g., via projective measurements. Here we employ the idea of space-time rotation of a circuit to explore the relation between systems that fall into these two classes. In particular, by space-time rotating unitary Floquet circuits  that display a localization transition, we construct non-unitary circuits  that display a rich variety of entanglement scaling and phase transitions. One outcome of our approach is a non-unitary circuit for free fermions in 1d that exhibits an entanglement transition from logarithmic scaling to volume-law scaling. This transition is accompanied by a `purification transition' analogous to that seen in hybrid projective-unitary  circuits. We follow a similar strategy to construct a non-unitary 2d Clifford circuit that shows a transition from area to volume-law entanglement scaling. Similarly, we space-time rotate a 1d spin chain that hosts many-body localization to obtain a non-unitary circuit that exhibits an entanglement transition. Finally, we introduce an unconventional correlator and argue that if a unitary circuit hosts a many-body localization transition, then the correlator is expected to be singular  in its non-unitary counterpart as well.
\end{abstract}

	\maketitle

\section{Introduction} \label{sec:intro}

Generic isolated quantum systems typically thermalize via the interaction between their constituents \cite{deutsch1991,srednicki1994chaos,srednicki1998, rigol2008, rigol_review}. One exception to this is the phenomenon of many-body localization (MBL) \cite{Basko_2006, Huse_2007_mbl,Huse_mbl_2010,huse_lbits,abanin_lbits, huse2015mbl,ros_lbits,altman2015mbl,imbrie2016,alet2018mbl,abanin2019review}where strong disorder causes the system to develop signatures of non-ergodicity such as sub-thermal entanglement under quantum quenches. More recently, it has been realized that new dynamical phases can emerge also in quantum systems subjected to projective measurements  \cite{Cao_Tilloy_2019, Skinner_2019, li2018measurment, Chan_2019, gullans2019dynamical, gullans2019scalable, zabalo2019critical, choi2019quantum, Tang_Zhu_2020, Li_2019, Szyniszewski_2019, zhang2020nonuniversal, goto2020measurementinduced, jian2019measurementinduced, bao2019theory, Skinner_2019, Li_2020, lang2020, nahum2020entanglement, sang2020measurement, Lavasani_2021, lavasani2021topological, Ippoliti_2021, duque2020topological, Turkeshi_2020, Fuji_2020, Lunt_2020, Szyniszewski_2020, vijay2020measurementdriven, Lopez_Piqueres_2020, Fidkowski_2021, nahum2021measurement, bao2021symmetry} due to the `quantum Zeno effect'   \cite{misra1977zeno}.  Relatedly, one can  consider  evolution with more general non-unitary circuits \cite{Chen_2020, tang2021quantum, jian2020criticality, Liu_2021, jian2021yanglee}, which typically exhibit non-ergodic behavior as well. It is natural to wonder if there is any relation between these two classes of systems, namely, unitarily evolved systems that show single-particle/many-body localization, and systems where non-unitarity plays a crucial role in suppressing ergodic behavior. In this work we explore such a connection using the idea of the space-time rotation of a circuit \cite{akila2016particle, Bertini2018PRL, Bertini2019PRX, chan2018spectral, chan2020spectral, Piroli2020, Kos2020, napp2020efficient, Ippoliti_2021b}.

For a unitarily evolved system to exhibit localization, spatial disorder of course plays a central role. Evidence suggests that time-translation invariance, whether continuous or discrete, is also crucial. For example, Floquet (i.e. time-periodic) circuits with spatial disorder can exhibit MBL phenomena \cite{Ponte_2015, Abanin_2016, Zhang_2016}, while unitary circuits that have randomness both in space and time tend to display ergodic behavior \cite{nahum_2017, khemani_operator_2017, nahum_operator_2018, rakovszky_diffusive_2017,von_keyserlingk_operator_2018, zhou_operator_2018, chen_power_law_2019, Zhou_Nahum_2019}. On the other hand, for the aforementioned non-unitary circuits displaying sub-thermal entanglement  \cite{Cao_Tilloy_2019, Skinner_2019, li2018measurment, Chan_2019, gullans2019dynamical, gullans2019scalable, zabalo2019critical, choi2019quantum, Tang_Zhu_2020, Li_2019, Szyniszewski_2019, zhang2020nonuniversal, goto2020measurementinduced, jian2019measurementinduced, bao2019theory, Skinner_2019, Li_2020, lang2020, nahum2020entanglement, sang2020measurement, Lavasani_2021, lavasani2021topological, Ippoliti_2021, duque2020topological, Turkeshi_2020, Fuji_2020, Lunt_2020, Szyniszewski_2020, vijay2020measurementdriven, Lopez_Piqueres_2020, Fidkowski_2021, nahum2021measurement, bao2021symmetry}, translation invariance in the time or the space direction is not crucial. This is demonstrated by explicit construction of circuits consisting of projective measurements dispersed randomly in space-time that host a transition from an area-law entanglement regime to a volume-law entanglement regime (see, e.g. Refs.\cite{Skinner_2019, li2018measurment, Chan_2019}). A sub-class of such non-unitary circuits have translation invariance in the space direction but lack translation invariance in the time direction. Such circuits will be the focus of this work for reasons we discuss next.

%

\begin{figure}[h]
	\centering
	\includegraphics[width=0.7\hsize]{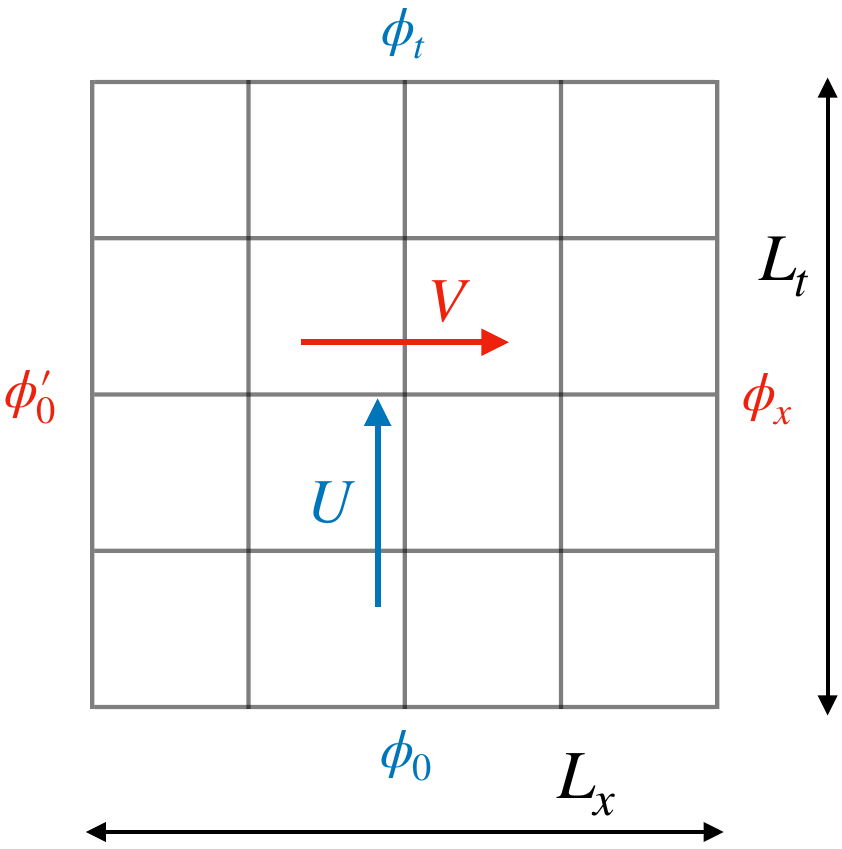}
	\caption{The geometry of circuit rotation employed in this work illustrated for a 1d system. Given a unitary circuit $U$ that acts on a system of spatial size $L_x$, the wavefunction evolved for time $L_t$ is schematically given by $\langle \phi_t|U|\phi_{0}\rangle = \int D\phi(x,t) e^{iS(\phi)}$ where the fields labeled $\phi'_{0}, \phi_x$ at the boundaries are also being integrated over in $D\phi(x,t)$ while $\phi_0, \phi_t$ act as boundary conditions. Using the same bulk action $S$, one may then define a rotated circuit $V$ that acts on a system with spatial extent $L_t$, such that the wavefunction at time $L_x$ is $\langle \phi_x|V|\phi'_{0}\rangle =  \int D\phi(x,t) e^{iS(\phi)}$.  In this rotated circuit, the fields labeled $\phi_{0}, \phi_t$ are being integrated over in $D\phi(x,t)$ while $\phi'_0, \phi_x$ act as boundary conditions.}
	\label{fig:circuitscheme}
\end{figure}

The main idea we will explore is the `space-time rotation' of a quantum circuit \cite{akila2016particle, Bertini2018PRL, Bertini2019PRX, chan2018spectral, chan2020spectral, Piroli2020, Kos2020, napp2020efficient, Ippoliti_2021b} with a focus on unitary circuits that host a localization-delocalization transition. To set the stage, consider a general unitary circuit $U$ that acts for time $L_t$ on a $d$-dimensional system of spatial size $L_1 \times L_2 \times ...\times L_d$. From this, one can define a `partition function' $Z = \tr(U)$. Denoting the underlying degrees of freedom schematically by symbol $\phi$, one may represent $Z$ as a path integral in space-time, $Z=\int D \phi(t,\{x\}) e^{ iS[\phi]}$, where $S[\phi(t,\{x\})]= \int dt d\vec{x} \,\mathcal{L}( \phi, t, \vec{x})$  is the space-time action and $\mathcal{L}(\phi, t, x_1, x_2, ...,x_d)$ is the corresponding Lagrangian. Let us now define a new Lagrangian $\tilde{\mathcal{L}}$ by interchanging $t$ and $x_1$: $\tilde{\mathcal{L}}(\phi, t, x_1, x_2,...,x_d) = \mathcal{L}(\phi, x_1, t, x_2, ...,x_d)$. For example, if $\mathcal{L}(\phi, t,x_1,x_2) = (\partial_t \phi)^2 - \left((\partial_{x_1} \phi)^4 + (\partial_{x_2} \phi)^6 + \phi^4\right)$, then $\tilde{\mathcal{L}} =  -(\partial_t \phi)^4 + (\partial_{x_1} \phi)^2 -\left( (\partial_{x_2} \phi)^6 + \phi^4\right)$. Since the original circuit $U$ is local, it implies that both $\mathcal{L}$ and $\tilde{\mathcal{L}}$ are also local. We use $\tilde{\mathcal{L}}$ to define a new `space-time rotated' circuit $V$: $\tr(V) = Z =  \int D \phi(t,\{x\}) e^{ i \int dt d\vec{x} \,\tilde{\mathcal{L}}}$. See Fig.\ref{fig:circuitscheme} for an illustration, and Sec.\ref{sec:spacetime} below for details. By design, the circuit $V$ acts for time $L_1$ on a system of spatial size $L_t \times L_2 ...\times L_d$.  Crucially, $V$ is \textit{not} guaranteed to be unitary  \cite{Bertini2018PRL}.  This point was recently employed in Ref.\cite{Ippoliti_2021b} to design a method for emulating certain non-unitary circuits and their associated measurement-induced phase transitions without requiring extensive post-selection. We note that in the context of \textit{imaginary} time evolution, the idea of space-time rotation to obtain a dual quantum Hamiltonians was first employed in Ref.\cite{betsuyaku1984study}.

In this work, we will perform the aforementioned space-time rotation on lattice models of Floquet circuits that are made out of unitaries with spatial disorder, and which display entanglement transitions due to the physics of localization. The rotated circuit $V$ will be generically non-unitary, and by construction, will possess translational invariance along a space direction, and disorder/randomness along the time direction. A motivation for our study is that the rotated and unrotated circuits have the same partition function $Z$, which is closely related to the spectral form factor \cite{guhr1998random, cotler2017black} $(= |Z|^2)$. Since the spectral form factor in a Hamiltonian/Floquet system is expected to show singular behavior across a localization transition \cite{Suntajs2020quantum, Prakash2021universal}, one may wonder if this fact has any consequence for the rotated circuit. In the special case when the rotation results in a unitary circuit, it was shown in Ref. \cite{Bertini2018PRL} that the (unrotated) Floquet circuit is chaotic. Here we instead start from Floquet circuits that can be argued to display a localization transition (and therefore not always chaotic), and study the non-unitary circuits that result from their rotation.

The first example we study corresponds to a Floquet circuit that displays an Anderson localization transition due to quasiperiodic disorder. Rotating this circuit results in a 1d \textit{free-fermion} non-unitary circuit that exhibits a transition from a volume-law entanglement regime, $S \sim L$ ($L$ is the spatial size), to a regime with entanglement characteristic of critical ground states: $S \sim \log(L)$. This is interesting because the known examples of non-unitary theories with free fermions have hitherto found only sub-extensive entanglement  \cite{Chen_2020, tang2021quantum, jian2020criticality, jian2021yanglee}. The fact that our non-unitary circuit is obtained from rotation of a unitary circuit  plays a crucial role in its ability to support volume-law entanglement.  


Next, we construct a 2d model where the unitary corresponds to a Floquet Clifford circuit and which displays a localization transition. Interestingly, space-time rotating this circuit results in a non-unitary circuit consisting only of unitaries and `forced' projective measurements. We find that both the rotated and the unrotated circuits display an entanglement phase transition from a volume law regime to an area-law regime.


The last example we study corresponds to a Floquet unitary circuit that displays an MBL transition  \cite{Ponte_2015, Abanin_2016, Zhang_2016}. The rotated, non-unitary counterpart again shows two distinct regimes, one where the entanglement scales as volume-law, and another where entanglement shows sub-extensive behavior.


Finally, we introduce an unconventional correlator that can be interpreted both within a unitary circuit and its non-unitary counterpart. We briefly discuss its measurement without employing any post-selection. Using the `$\ell$-bit' picture of MBL \cite{huse_lbits,abanin_lbits}, we provide a heuristic argument that this correlator  exhibits singular behavior across an MBL transition.

The paper is organized as follows. In Sec.\ref{sec:spacetime}, we provide a brief overview of the idea of space-time rotating a circuit. In Sec.\ref{sec:1d_floquet_ising}, we discuss a Floquet model of non-interacting fermions in 1d that displays a localization-delocalization transition due to quasiperiodicity. We then study the phase diagram of the non-unitary circuit that results from its space-time rotation. In Sec.\ref{sec:2d_floquet_clifford} we discuss a two dimensional Clifford Floquet circuit that displays a localization-delocalization transition, and then study its space-time rotated version that turns out to be a hybrid circuit consisting of only unitaries and forced projective measurements. In Sec.\ref{sec:floquet_mbl}, we discuss a 1d interacting Floquet model that displays many-body localization transition and study the phase diagram of its rotated counterpart. In Sec.\ref{sec:correlations} we introduce an unconventional correlator and discuss its physical consequences.  Finally, in Sec.\ref{sec:discuss}, we conclude with a discussion of our results.

\section{Brief overview of Space-time rotation of a circuit} \label{sec:spacetime}

Here we briefly review the idea of the space-time rotation of a circuit using a 1d lattice model \cite{Bertini2018PRL}. Although we specialize to 1d for now, the discussion can be straightforwardly generalized to higher dimensions, as we do so in Sec.\ref{sec:2d_floquet_clifford}. We begin by considering the following unitary Floquet circuit for a system of spatial size $L_x$:

\begin{equation}\label{eq:main_floquet}
U_F=  e^{ i   \sum_{r=1}^{L_x} J_{X,r}X_r   }  e^{  i  \sum_{r=1}^{L_x} J_{Z,r} Z_rZ_{r+1}  + i   \sum_{r=1}^{L_x} h_r Z_r }.
\end{equation}
As discussed in the introduction, a space-time rotated mapping is constructed by investigating the `partition function' $Z =  \tr \left[   (U_F)^{L_t} \right]$. Using the standard quantum-classical mapping,
$Z$ can be expressed as a partition function of $L_x \times L_t$ number of classical variables $\{s_{r,t}\}$ in two dimensions with complex Gibbs weight : $Z \propto  \sum_{ \{  s_{r,t} \}   } e^{ -S}$, where the action $S$ reads
\begin{equation}
\quad -S= \sum_{r,t}  \left( i\tilde{J}_{Z,r}  s_{r,t} s_{r,t+1} + iJ_{Z,r}  s_{r,t} s_{r+1,t} +  ih_rs_{r,t} \right).
\end{equation}
The coupling between neighboring spins $\tilde{J}_{Z,r} s_{r,t} s_{r,t+1} $ along  the time direction results from  $e^{ i  J_{X,r} X_r  }$ in the Floquet unitary $U_F$, and the coupling constant $\tilde{J}_{Z,r}$ is determined as $\tilde{J}_{Z,r} = - \pi /4 +  \frac{i}{2}  \log\left(   \tan J_{X,r}   \right)$. To obtain the space-time rotated circuit, one can now define a Hilbert space for $L_t$ number of spins on a given fixed-$r$ time-like slice (Fig.\ref{fig:circuitscheme}). Correspondingly, the partition function can be written as $Z \propto \tr \left[     \prod_{r=1}^{L_x} V_r  \right]$, where $V_r $ acts on a Hilbert space of $L_t$ spins: $ V_r=  e^{  i  \tilde{J}_{X,r} \sum_t X_t  } e^{  i   \tilde{J}_{Z,r}  \sum_t Z_tZ_{t+1} +i h_r \sum_t  Z_t  }$ with $\tilde{J}_{X,r}=  \tan^{-1}\left( -i e^{ -2i J_{Z,r}}  \right) $. Altogether, $e^{i J_{X,r}X_r}$ in $U_F$ is mapped to $e^{i\tilde{J}_{Z,r}  Z_t Z_{t+1} }$  in $V_r$, and $e^{i J_{Z,r} Z_rZ_{r+1}}$ in $U_F$ is mapped to $e^{i\tilde{J}_{X,r} X_t }$ in $V_r$.

Finally, by exchanging the labels of space-time coordinates $r \leftrightarrow t$, one can  construct the  space-time rotated circuit $V(T)$ that evolves the system for a time $T$ and acts on a Hilbert space of size $L$, $V(T) = \prod_{t=1}^T V_t$ where
\begin{equation} \label{eq:defV}
V_t=  e^{  i  \tilde{J}_{X}(t)  \sum_{r=1}^{L} X_r  } e^{  i   \tilde{J}_{Z}(t)   \sum_{r=1}^{L} Z_rZ_{r+1} +i h(t) \sum_{r=1}^{L}   Z_r  }.
\end{equation} 
A few remarks are in order. First,  $V(T)$ has the space translational invariance resulting  from the time translation invariance in the unrotated Floquet circuit $U_F$. Second, $V(T)$ is generically non-unitary except for the self-dual points $\abs{J_{X,r}}=\abs{J_{Z,r}}=\pi/4$ \cite{Bertini2018PRL}. Third, in the special case when  $J_{X,r}$, $J_{Z,r}$, and $h_{r}$ are restricted to $\{0, \pm \pi/4 \}$, $V(T)$ corresponds to a hybrid quantum circuit with only unitary gates and forced projective measurements. While a $\pi/4$ coupling gives unitary operation as just  mentioned, $J_{X,r}=0$ implies that  the spin at site $r$ is frozen in the unrotated circuit, and hence, in the rotated circuit, this corresponds to a forced projective measurement of $(1+Z_iZ_{i+1})/2$ on two neighboring spins. Similarly, $J_{z,r}=0$ corresponds to a forced projective measurement of $(1+X)/2$ on a single site. The fact that a forced projective measurement can arise from the space-time rotation of a unitary gate has also been previously noted in Ref.\cite{Ippoliti_2021b}.  Finally, once we have obtained the form of $V$, we let the corresponding system size $L$ and the evolution time $T$ (Eq.\ref{eq:defV}) be free parameters that are independent of the system size and evolution time of the Floquet unitary $U_F$ from which it was obtained. That is, we \textit{do not} impose the conditions $T = L_x, L = L_t$, when we compare various properties of $V$ with $U_F$.

Having reviewed the mapping between a unitary and its `space-time dual', in the rest of the paper we will consider several Floquet unitary circuits that exhibit entanglement transitions due to the physics of localization, and explore the phase diagrams of their space-time duals.

\section{Space-time rotation \& Entanglement transition in a quasiperiodic circuit}\label{sec:1d_floquet_ising}

\begin{figure*}[ht]
	\centering
	\begin{subfigure}[b]{0.243\textwidth}
		\includegraphics[width=\textwidth]{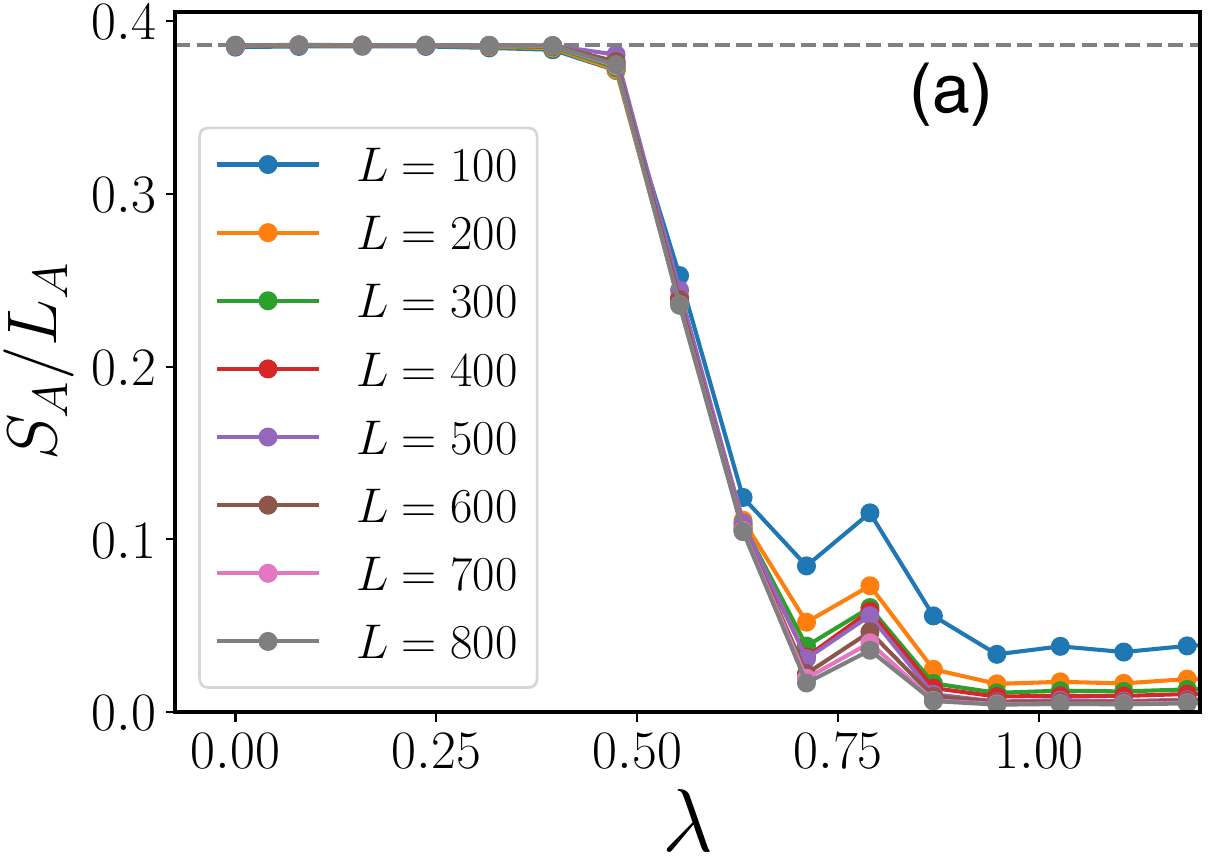}
	\end{subfigure}
	\begin{subfigure}[b]{0.243\textwidth}
		\includegraphics[width=\textwidth]{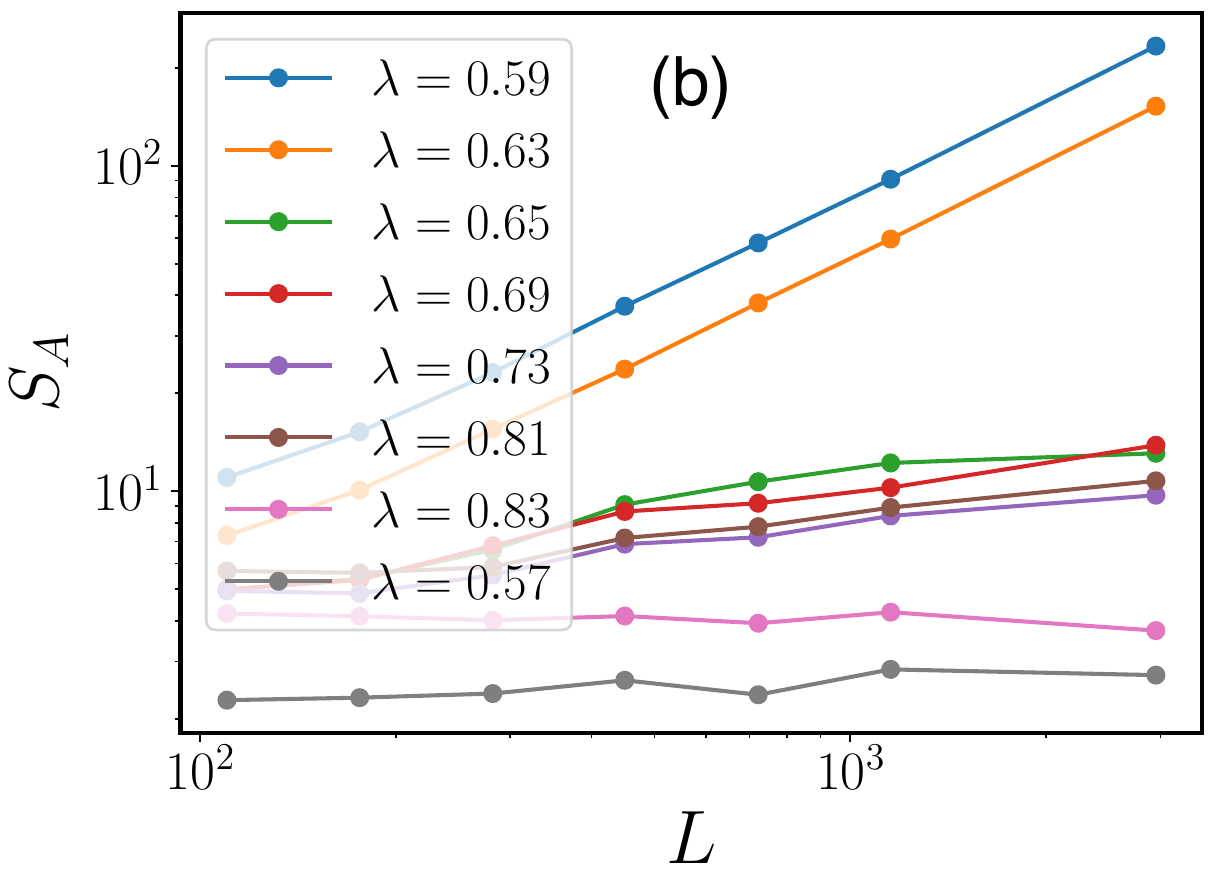}
	\end{subfigure}	
	\begin{subfigure}[b]{0.243\textwidth}
		\includegraphics[width=\textwidth]{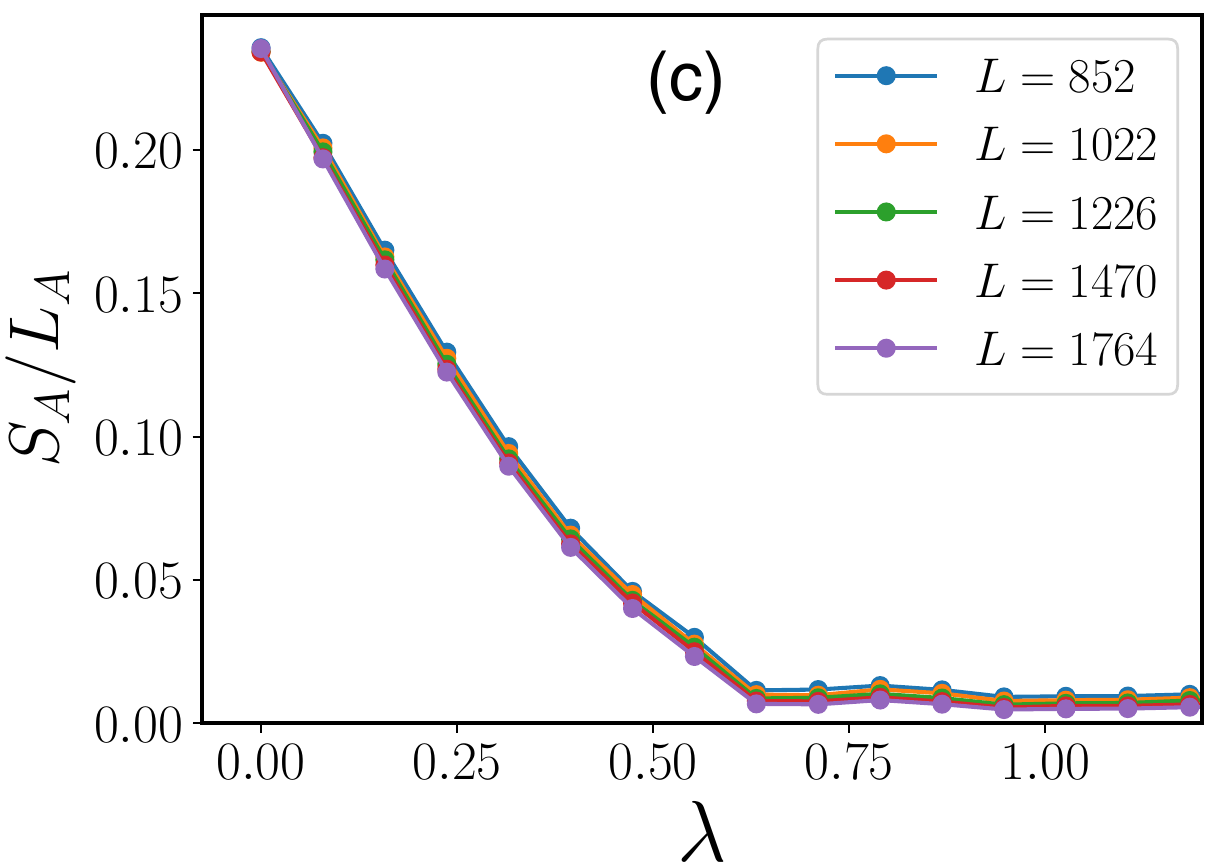}
	\end{subfigure}	
	\begin{subfigure}[b]{0.243\textwidth}
		\includegraphics[width=\textwidth]{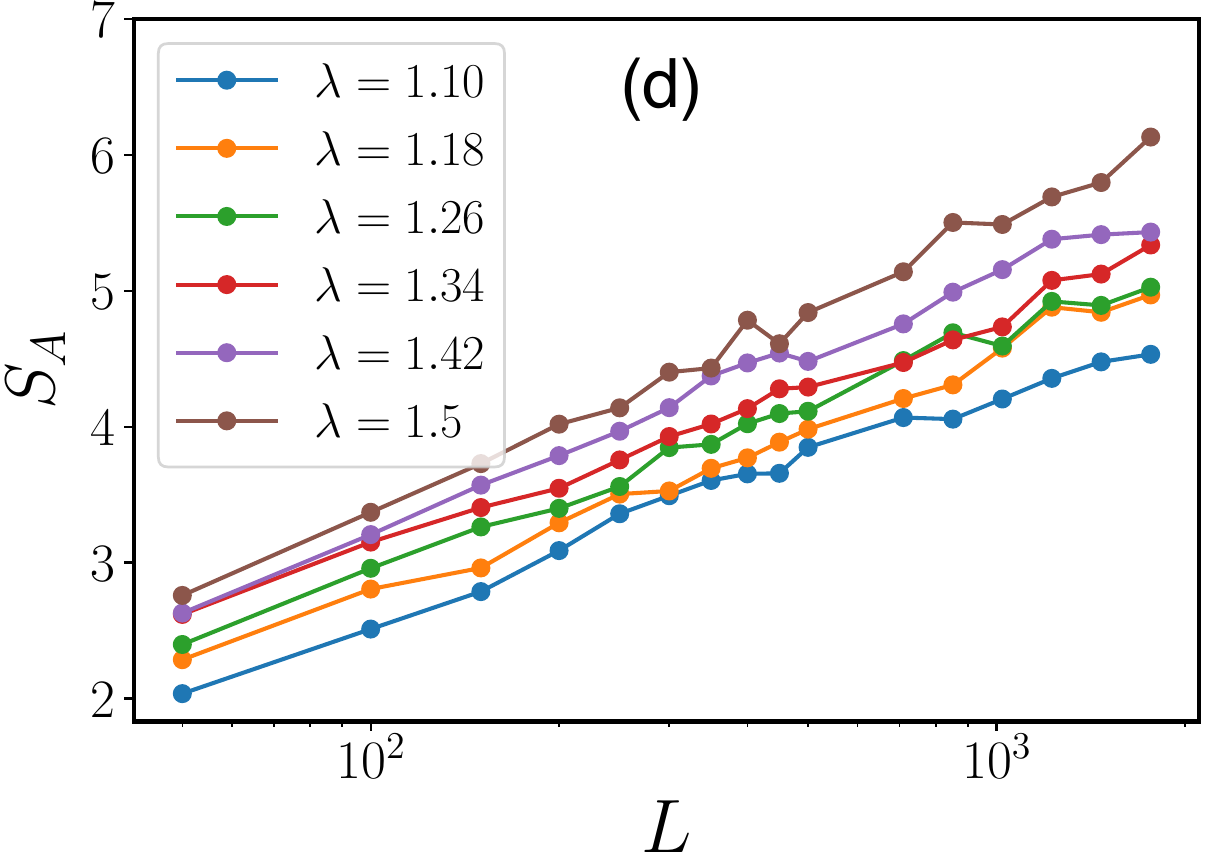}
	\end{subfigure}
	\caption{Entanglement entropy $S_A$ of long-time-evolved states (evolution time up to $O(L^2)$) at the subsystem size fraction $L_A/L=1/2$. (a) Entanglement for the Floquet circuit, Eq.\ref{eq:AA_floquet}. The dashed gray line marks the entanglement entropy density averaged over all eigenstates of random quadratic Hamiltonians \cite{Rigol_2020_random_quadratic}.  (b) Scaling of $S_A$ with $L$ in a narrower range of $\lambda$ for the same Floquet circuit as Fig.(a). $S_A$ scales linearly with $L$ (i.e. volume-law) for $\lambda \lesssim 0.64$ and follows an area law for $\lambda \gtrsim 0.81$. For $0.64 \lesssim \lambda \lesssim 0.81$, $S_A\sim O(L^{\gamma})$ with $0< \gamma<1$.  (c) Entanglement entropy density $S_A/L_A$ for the space-time-rotated non-unitary circuit in Eq.(\ref{eq:1d_rotated_circuit}). $S_A/L_A$ is non-zero for small modulation strength $\lambda$, while it vanishes for large $\lambda$. (d) Entanglement scaling in the regime $1.5 \gtrsim \lambda \gtrsim 1$ for the same circuit as Fig.(c). One finds $S_A \sim \log(L)$.} 
	\label{fig:1d_longtime_s}
\end{figure*}

As a first example, we consider a Floquet circuit in one space dimension hosting a localization-delocalization transition. We recall that models with quasiperiodic randomness, such as the Aubry-Andr\'{e}-Harper (AAH) model\cite{harper1955general, azbel1979quantum, aubry1980}, can evade Anderson localization \cite{anderson1958} in 1d. The AAH model is given by $H= -t \sum_{r} \left(c_r^{\dagger}c_{r+1} + h.c. \right) -  2\lambda\sum_r \cos(2\pi Q r +\delta)c_r^{\dagger}c_r$, where $c_r$ and $c_r^{\dagger}$ are the fermion creation and annihilation operators. When the on-site potential is incommensurate, i.e., the wavenumber $Q$ is irrational, all single-particle eigenstates are delocalized (localized) for $\abs{t}> \abs{\lambda}$ ($\abs{t}< \abs{\lambda}$) and arbitrary offset $\delta$. Motivated by this, we consider a Floquet circuit model with the following unitary 

\begin{equation}\label{eq:AA_floquet}
U_F =e^{ i J \sum_r  X_rX_{r+1}  }e^{ i  \sum_r h_rZ_r}
\end{equation}
for a spin-1/2 chain of size $L$ with periodic boundary conditions. We choose $J=1$, and $h_r$ to be quasiperiodic: $h_r=h+ \lambda \cos( 2\pi Qr +\delta )$ where $Q$ is set to $\frac{2}{1+\sqrt{5}}$ (the inverse Golden ratio), and $h = 2.5$. We note that Ref.\cite{Laumann2017} studied the incommensurate AAH modulation in the transverse field Ising model, and found that due to the interplay between symmetry and incommensurate modulation, it exhibits a rich phase diagram, including phases with delocalized, localized, and critical states that sometimes also break the Ising symmetry spontaneously.  

Using the above Floquet unitary $U_F$, we construct the corresponding space-time-rotated circuit $V$ as discussed above in Sec.\ref{sec:spacetime}:
\begin{equation}\label{eq:1d_rotated_circuit}
V(T)= \prod_{t=1}^T e^{ i  \tilde{h}\sum_r  Z_r  }  e^{ i  \sum_r  \tilde{J  }(t) X_r X_{r+1}  }.  
\end{equation}
where $\tilde{J}(a) = - \pi /4 +  \frac{i}{2}  \log\left( \tan h_a  \right)$ and $\tilde{h}=  \tan^{-1}\left( -i e^{ -2i J}  \right) $.
Notice that the circuit $V$ is translationally invariant in space at each fixed time slice, but quasiperiodic in time.

Now we discuss the entanglement structure of long-time-evolved states ($T\gg L$) from a product state $\ket{\psi_0}$: 
\begin{equation}
\ket{\psi(T)} =  \frac{  U \ket{\psi_0}   }{  \sqrt{  \bra{ \psi_0}  U^{\dagger}    U \ket{\psi_0} }}, 
\end{equation}
where $U$ is chosen as $(U_F)^T$ and $V(T)$ for the Floquet circuit and its space-time dual respectively. Using the Jordan-Wigner transformation, we map these circuits into a problem involving free-fermions, and numerically compute the entanglement entropy using the correlation matrix technique \cite{Chung2001density, cheong2004manybody, Peschel_2003}(see Appendix.\ref{appendix:AA_ising_entanglement} for the details).

For the unrotated circuit $U_F$, we find that the entanglement entropy exhibits a volume-law scaling for  $\lambda \lesssim 0.64 $ and an area-law scaling for $\lambda \gtrsim 0.81$ (Fig.\ref{fig:1d_longtime_s}(a)). In the intermediate regime, $0.64 \lesssim \lambda \lesssim 0.81$ (Fig.\ref{fig:1d_longtime_s}(b))  we find that $S_A\sim O(L^{\gamma})$ with $0< \gamma<1$. Notably, deep in the volume-law phase, the entanglement entropy density $S_A/L_A\approx 0.386$ regardless of $\lambda$, which is very close to the average value predicted for random quadratic Hamiltonians of free fermions derived in Ref.\cite{Rigol_2020_random_quadratic}: $s_{r} = \log 2 - \left[  1+ f^{-1} (1-f) \log (1-f )  \right] \approx 0.386$ at $ f = 1/2$. We also explore delocalization properties of the single-particle eigenfuntions of the circuit $U_F$ in terms of free fermions and find three distinct phases (see Appendix.\ref{appendix:AA_single_particle}), in line with the late-time entanglement entropy studied here. 

We now discuss the space-time-rotated circuit $V$. We find that it also exhibits a transition in the entanglement entropy of long-time evolved states.  Fig.\ref{fig:1d_longtime_s}(c) indicates that there is a transition in the entanglement entropy density $S_A/L_A$ at $\lambda \approx 0.64$: $S_A$ follows a volume law for $\lambda \lesssim 0.64$, and obeys a sub-volume scaling for $\lambda \gtrsim 0.64$.  We also note that in the volume-law regime, the coefficient of the volume law varies continuously, in strong contrast to the volume-law phase of the unrotated unitary circuit. To elucidate the nature of the sub-volume-law regime, we study $S_A$ Vs $L$, and find that it scales logarithmically with the system size $L$: $S \sim \alpha \log(L)$ (see Fig.\ref{fig:1d_longtime_s}(d)) where $\alpha$ is a number that depends on $\lambda$. We also attempted a scaling collapse for the entanglement close to the critical point in the non-unitary circuit, see Appendix \ref{appendix:collapse}. The collapse is reasonably good in the volume-law regime while it doesn't work  well in the sub-volume-law regime. We suspect that this may be  related to the fact that the coefficient $\alpha$ in the logarithmic scaling of entanglement varies continuously with $\lambda$.

A heuristic argument relates the physics of localization in the unitary circuit to the physics of quantum Zeno effect \cite{misra1977zeno} in the rotated non-unitary circuit, and also suggests that the aforementioned entanglement transition is likely to occur at $\lambda_c = \pi -h\approx 0.64$, in line with our numerical observations. For the unrotated circuit $U_F$, the condition $\lambda>\lambda_c$ implies that some $h_r$ in the term $e^{ i  h_rZ_r}$ is arbitrary close to $\pi$. Mapping the spin chain to Majorana fermions using the Jordan-Wigner transition, the corresponding location $r$ then has a broken bond between two neighboring sites of the Majorana fermions, thereby impeding their propagation. In contrast, from the point of view of the rotated circuit $V$,  the space-time rotation of the term $e^{ih_r Z_r}$ at $h_r= \pi$ corresponds to the two-spin gate $e^{i\tilde{J} X_jX_{j+1}  }$ with $\tilde{J}= - \pi/4 + \frac{i}{2}  \log\tan(\pi)$, which therefore acts as a projector $\frac{1}{2}(1+X_jX_{j+1})$. Crucially, such a two-site projection occurs uniformly in space (due to the space translational symmetry of the rotated circuit), leading to the absence of volume-law entanglement for time-evolved states.

Perhaps the most surprising aspect of our result is the presence of a  volume-law phase since the previous works on non-unitary free-fermion circuits found phases only with sub-extensive entanglement  \cite{Chen_2020, tang2021quantum, jian2020criticality, jian2021yanglee}. For hybrid circuits consisting of unitary evolution interspersed with projective measurements, it was found in Ref.\cite{Cao_Tilloy_2019} that volume-law entanglement in a free-fermion chain is destroyed by the presence of arbitrarily weak measurement. Ref.\cite{Fidkowski_2021} argued for similar results. However, these results do not contradict ours since in the volume-law phase, our non-unitary circuit does not specifically correspond to unitary evolution interspersed with projective measurements but instead corresponds to more general evolution with a non-Hermitian Hamiltonian (see Eq.\ref{eq:1d_rotated_circuit}).

To gain intuition for the origin of the volume-law phase, we consider a simplified circuit that has translation symmetry in both space and time: $ V_0= e^{  i J\sum_jX_jX_{j+1}} e^{   i h \sum_j Z_j}$, and allow $J$ and $h$ to be  complex numbers. If $V_0$ is obtained from the space-time rotation of a unitary circuit, a key feature is that the real part of both $J$ and $h$ will be $\pi/4$. Writing $J= \pi /4 + i \alpha_J$ and $h= \pi /4 + i\alpha_h$, we find analytically that such a circuit leads to volume-law entanglement at long times for any $\alpha_J$ and $\alpha_h$ (see Appendix.\ref{appendix:pi_over_four}).  The volume-law phase originates from the fact that when $\textrm{Re}(J)=\textrm{Re}(h)= \pi/4$, an extensive number of single-particle eigenvalues of the Floquet unitary are real. Setting $\alpha_J=\alpha_h= \alpha$, and using a simple quasiparticle picture\cite{cardy_quench_2005}, we find that the volume-law coefficient of entanglement decays exponentially with $\alpha$: $S_A/L_A\sim e^{-c \alpha}$ for $c>0$. Therefore, there is no area-law phase in this simplified, translationally invariant model. We numerically verified these results as well. Although we don't have similar analytical results for the circuit $V$ (Eq.\ref{eq:1d_rotated_circuit}), we verified numerically that Re$(\tilde{J})=$ Re$(\tilde{h}) = \pi/4$ (due to the circuit being obtained from the rotation of a unitary, namely $U_F$) is again essential to obtain a volume-law phase. In this sense,  the volume-law phase of the non-unitary circuit is `symmetry-protected' by the unitarity of the unrotated circuit.

One may also inquire about the role played by the time-translation symmetry of the unitary circuit. If one chooses a different unitary circuit for each time slice, then the localization is lost at any $\lambda$ and one only obtains a volume-law phase in the corresponding unitary circuit. We verified that the rotated circuit, which now lacks spatial translational symmetry, does not exhibit a phase transition. Therefore, at least for this specific problem, both the unitarity and the translation symmetry plays a crucial role to obtain the entanglement transition.

\begin{figure}[h]
	\centering
	\begin{subfigure}[b]{0.45\textwidth}
		\includegraphics[width=\textwidth]{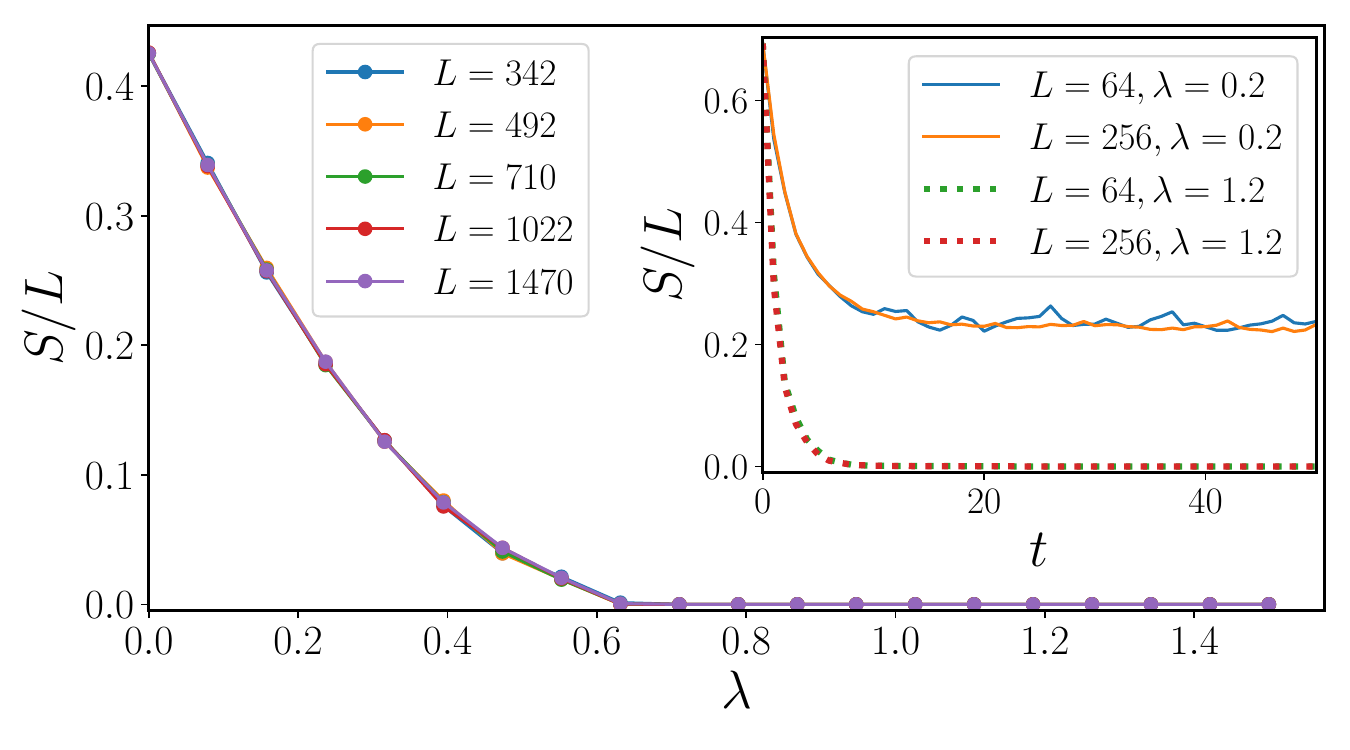}
	\end{subfigure}
	\caption{The late-time entropy density $S/L$ for a density matrix that is initially in a completely mixed state (i.e. $\rho(t=0) \propto \mathbb{1}$) and is evolved with the non-unitary circuit defined in Eq.\ref{eq:1d_rotated_circuit}. Inset: Time evolution of $S/L$.} \label{fig:1d_purification}
\end{figure}

It was argued in Refs. \cite{gullans2019dynamical, gullans2019scalable} that a stable volume-law entangled phase of pure states in a hybrid unitary-projective circuit is a consequence of the robust error-correcting properties of the circuit against environmental monitoring. Consequently, a maximally mixed state $\rho  = \mathbb{1}/2^L$ evolved by the circuit will retain a finite residual entropy density up to an extremely long time, indicating stability against purification by monitoring. Motivated by these results, we studied the purification dynamics of a maximally mixed state evolved under our non-unitary circuit and investigated its von Neumann entropy density as a function of time. Remarkably, we find a sharp transition in the entropy density, where for $\lambda <  0.64$ (i.e. the volume-law entanglement phase), the system has a non-zero entropy density even at times $t \gg L$, and for $\lambda> 0.64$, the system is purified with a vanishing entropy density in a time that is independent of the system size $L$ (see Fig.\ref{fig:1d_purification} and Appendix.\ref{appendix:AA_purification}).

\section{Space-time rotation \& Entanglement transition in a 2d Clifford circuit}\label{sec:2d_floquet_clifford}

\begin{figure*}[ht]
	\centering
	\begin{subfigure}[b]{0.243\textwidth}
		\includegraphics[width=\textwidth]{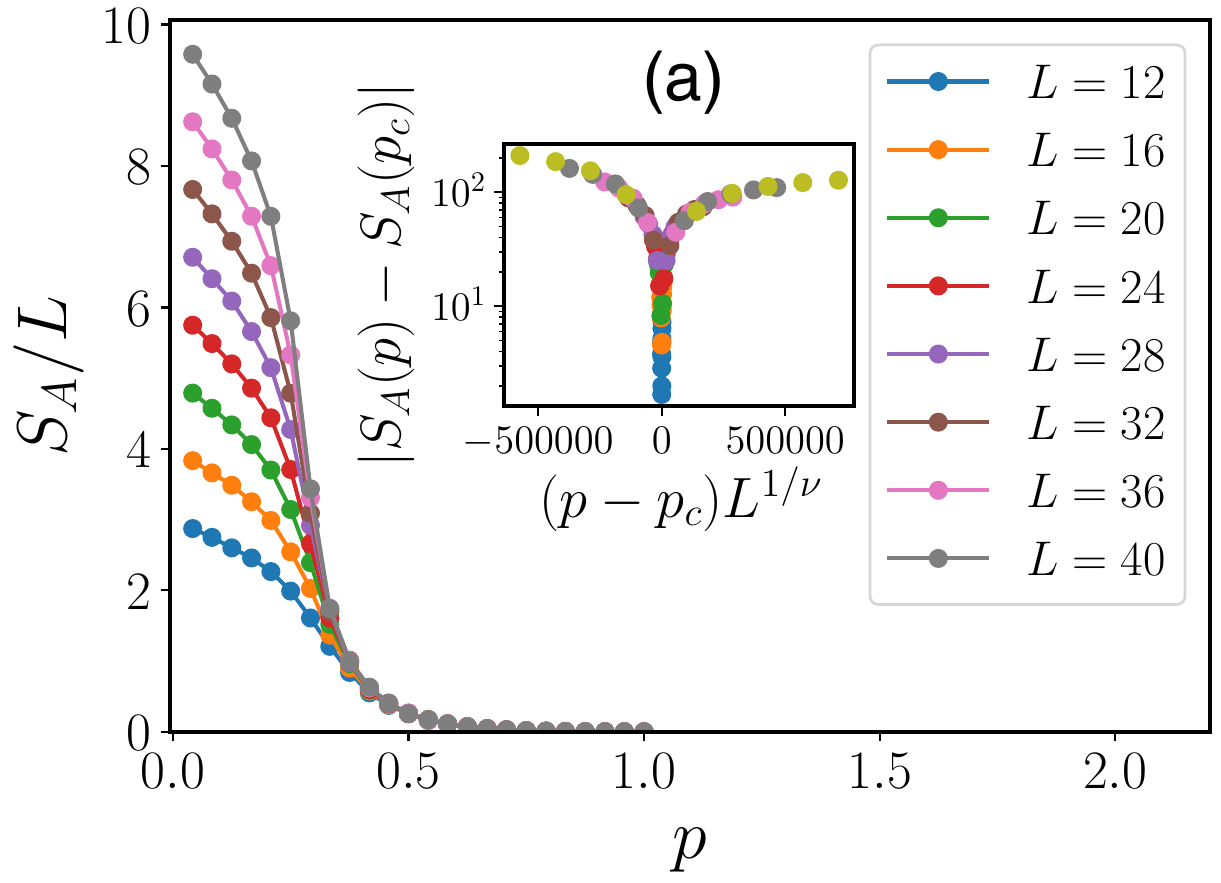}
	\end{subfigure}
	\begin{subfigure}[b]{0.243\textwidth}
		\includegraphics[width=\textwidth]{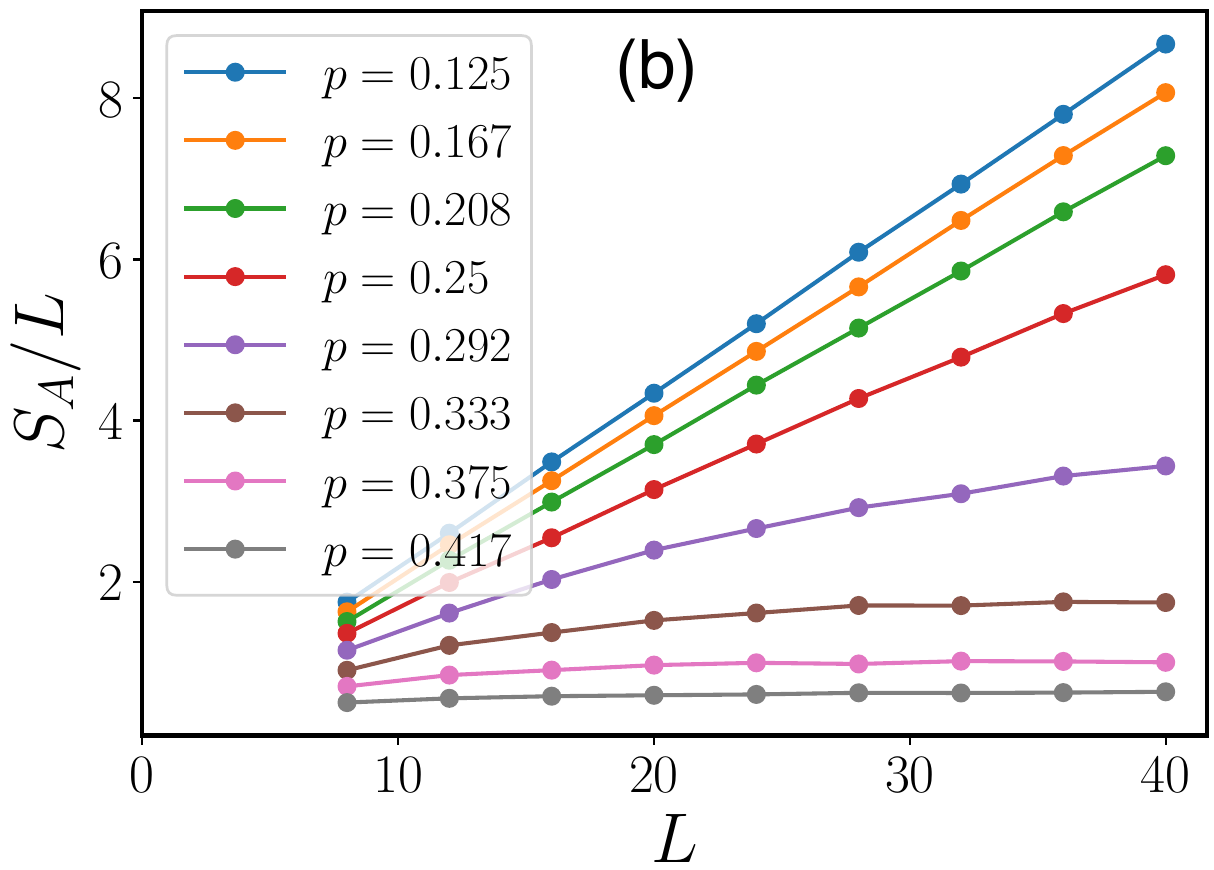}
	\end{subfigure}
	\begin{subfigure}[b]{0.243\textwidth}
		\includegraphics[width=\textwidth]{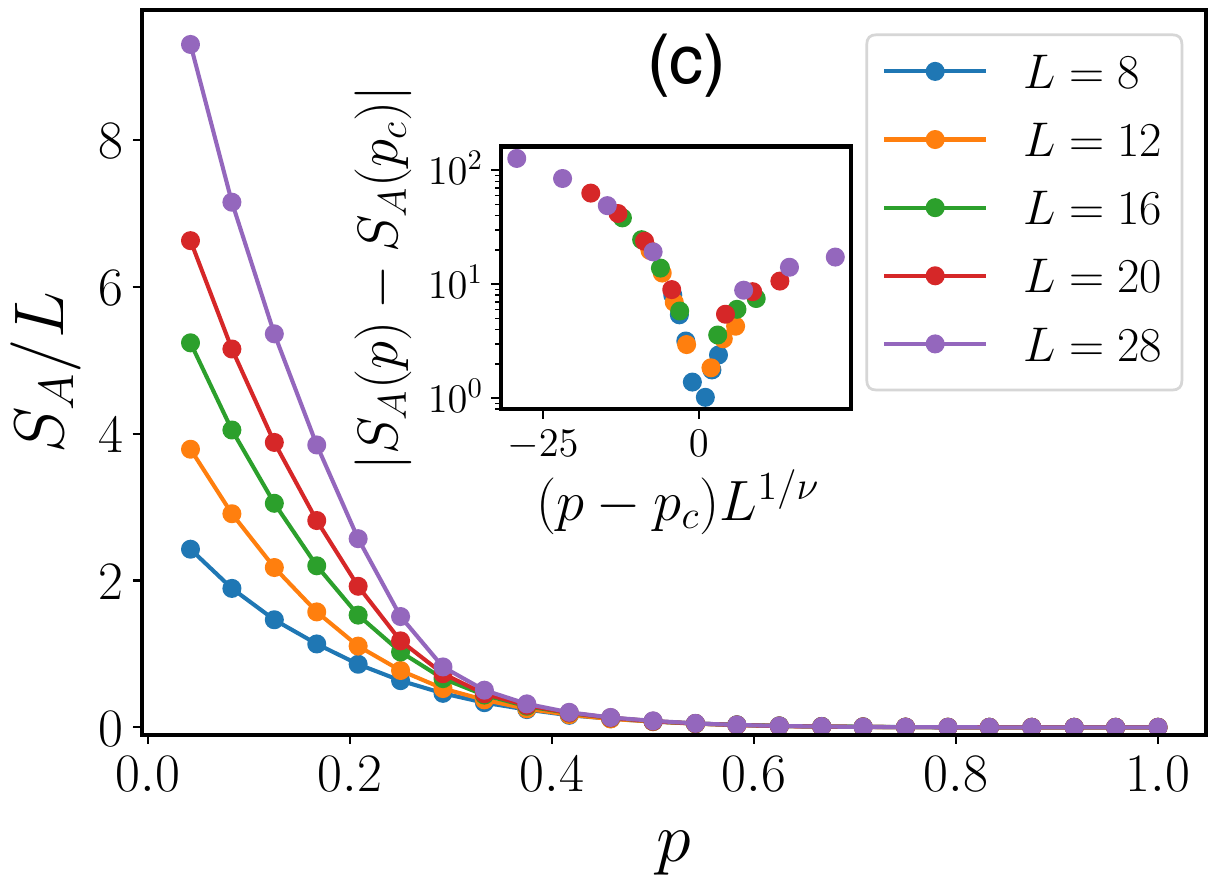}
	\end{subfigure}
	\begin{subfigure}[b]{0.243\textwidth}
		\includegraphics[width=\textwidth]{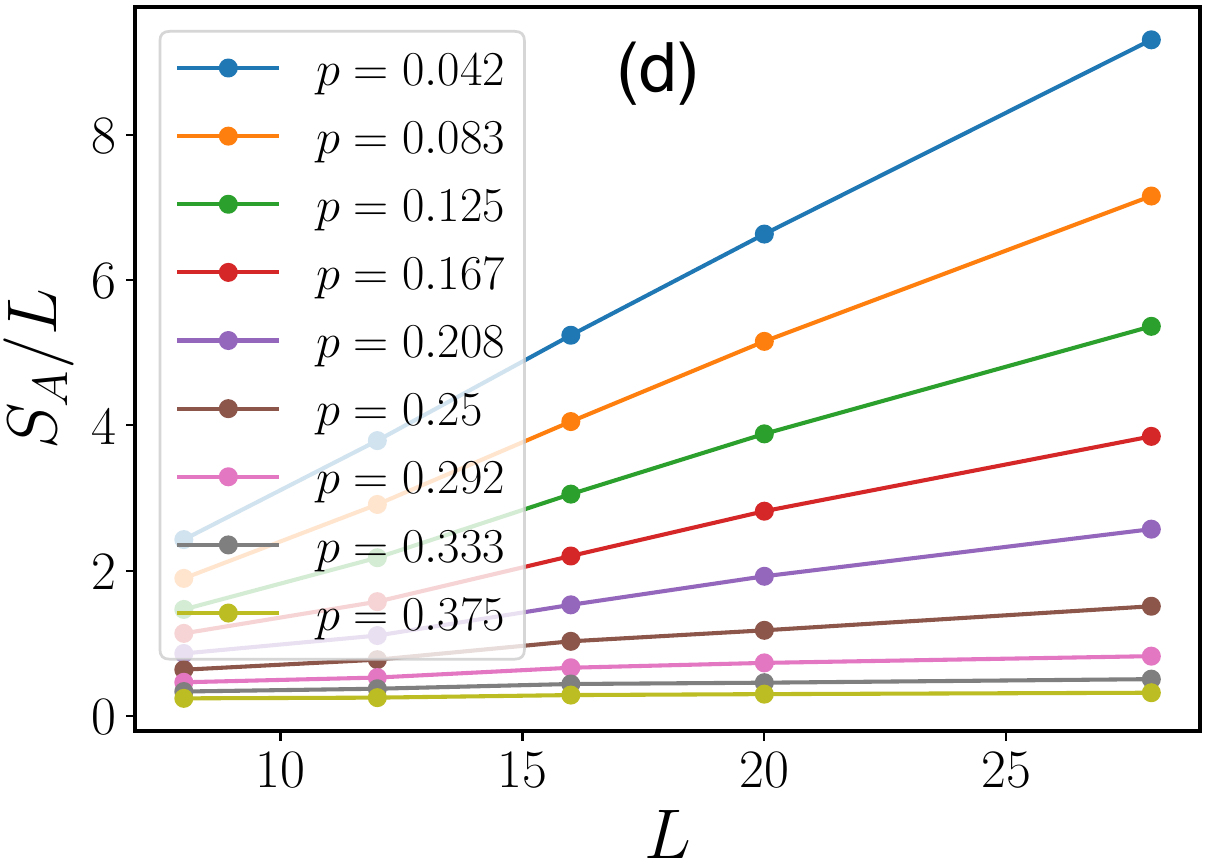}
	\end{subfigure}
	\caption{(a) Long-time entanglement entropy $S_A$ of a subregion $A$ of size $L/2\cross L$ averaged over $O(10^3)$ random realizations of the unitary circuit defined in Eq.\ref{eq:2dim} as a function of the probability $p$. The inset shows the scaling collapse across the critical point between a volume-law regime and an area-law regime with  $(p_c, \nu) \approx (0.28,0.38)$. (b) $S_A/L$ for the same system as Fig.(a) as a function of $L$ for various values of $p$. For small $p$, $S_A/L \sim L$ i.e. $S_A \sim L^2$, while for large values of $p$, $S_A/L$ is independent of $L$ signifying area-law scaling. (c) Long-time entanglement entropy of a subregion of size $L/2\cross L$ averaged over $O(10^4)$ random realizations of the rotated non-unitary circuit (Eq.\ref{eq:2dim_rotated}). The inset shows the scaling collapse with $ (p_c, \nu) \approx (0.28,0.49)$. (d) $S_A/L$ for the same system as Fig.(c) as a function of $L$ for various values of $p$. Again, for small $p$, $S_A/L \sim L$ i.e. $S_A \sim L^2$, while for large values of $p$, $S_A/L$ is independent of $L$.}	
	\label{fig:2dim}
\end{figure*}

We next explore entanglement transitions in a two-dimensional Floquet model and its space-time dual. We consider the following Floquet unitary on a square lattice of size $L \times L$:
\begin{equation}\label{eq:2dim}
U_F = e^{ -i \frac{\pi}{4} \sum_{\expval{ij}} J_{ij}  Z_i Z_j   } e^{  -i\frac{\pi}{4} \sum_i h_i X_i },
\end{equation} 
Here each $J_{ij}$, $h_i $ is chosen to be $0$ or $1$ with probability $p$ and $1-p$ respectively. This is a Clifford circuit since it maps a Pauli string to another Pauli string: $e^{i  \frac{\pi}{4} Z_1Z_2  } X_j e^{- i  \frac{\pi}{4} Z_1Z_2  } = i Z_1   Z_2X_j$ for $j=1,2$ and $e^{i  \frac{\pi}{4} X_j } Z_j e^{- i  \frac{\pi}{4} X_j  } =  iX_j Z_j$. Therefore it can be efficiently simulated based on the Gottesman-Knill theorem\cite{gottesman_1996,gottesman1998heisenberg,gottesman_2004}. The construction of the circuit $U_F$ is motivated from Ref.\cite{Chandran2015semiclassical}, although it differs from the precise circuit discussed in that work.

To construct the space-time-rotated circuit, we interchange the time coordinate $t$ and one of the space coordinates $x$ while leaving the other space coordinate $y$ unchanged. This results in the mapping as follows. Since $y$ coordinates are unchanged, the gate $e^{ -i \frac{\pi}{4} J_{ij} Z_iZ_j    }$, with $\expval{ij}$ being a $y$-directed bond, is invariant under the space-time-rotated mapping. The gate $e^{ -i \frac{\pi}{4} J_{ij}  Z_iZ_j    } $  along the $x$ direction in the unrotated circuit is mapped to a single-site gate $ e^{ i \frac{\pi}{4} X } , \frac{1+ X}{2} $ in the rotated circuit for $J_{ij} =1, 0$ respectively. Finally, the single-site gate $e^{ -i \frac{\pi}{4}  h_iX   } $ in the unitary circuit maps to the two-site gate  $  e^{ i \frac{\pi}{4} Z_iZ_j } , \frac{1+ Z_iZ_j}{2}  $ on an $x$-directed bond in the non-unitary circuit for $  h_{i} =1, 0$ respectively. Therefore, the rotated circuit consists of unitary evolution interspersed with forced projective measurements, and is given by

\begin{equation}\label{eq:2dim_rotated}
V(T)= \prod_{t=1}^T\left[  \prod_{y=1}^{L_y}\left[   V_X(y,t) V_{ZZ,|}(y,t) V_{ZZ,- }(y,t)       \right]    \right]
\end{equation}
where for each $y$ and $t$, $V_{X}(y,t)    =  \prod_{x=1}^{L_x}   \frac{1+X_{x,y}   }{2} $ or $ \prod_{x=1}^{L_x} e^{ i \frac{\pi}{4}  X_{x,y} }$ with probability $p $ and $1-p$,  $V_{ZZ,|}(y,t) = 1$ or $  \prod_{x=1}^{L_x} e^{ i \frac{\pi}{4}  Z_{ x,y}  Z_{x,y+1} } $ with probability $p$ and $ 1-p$, and $V_{ZZ,-}(y)   =    \prod_{x=1}^{L_x}   \frac{1+Z_{x,y}Z_{x+1,y}    }{2}$ or $  \prod_{x=1}^{L_x} e^{ i \frac{\pi}{4}  Z_{ x,y}  Z_{x+1,y} } $ with probability $p$ and $ 1-p$. Note that $V$ has translation symmetry along $x$ inherited from the time translation symmetry in the unrotated Floquet circuit.

Now we discuss the entanglement structure of long-time-evolved states.  For both unrotated and rotated circuit,  we find an entanglement transition between a volume-law phase and an area-law phase at the same finite critical probability $p=p_c \approx 0.28$ (see Fig.\ref{fig:2dim}). Assuming the following scaling form of entanglement entropy $\abs{   S_A(p)-S_A(p_c)  } = F((p-p_c)L^{1/\nu} ) $, we find that the correlation length exponent $\nu$ however differs in the two circuits ($\nu\approx 0.38$ for the unrotated circuit and $\nu\approx 0.49$ for the rotated one). The coefficient of the volume-law entanglement varies continuously in both circuits and vanishes continuously across the phase transition.

We also analyzed entanglement scaling at the critical point, and found that both in the rotated and the unrotated circuit, the data is indicative of the scaling $S \sim L \log L$, which is reminiscent of results in Refs.\cite{nahum2020entanglement, Turkeshi_2020, tang2021quantum}, see Appendix \ref{appendix:2dlogscaling}. However, as pointed out in Ref.\cite{Lunt_2020}, on small system sizes, a slight error in the location of the critical point can make an area-law scaling, $S \sim L$, appear as $S \sim L \log L$ scaling. Therefore, one may need to study larger system sizes to be conclusive. As an aside, we note that the scaling form $S \sim L \log L$ is not allowed for a system described by a unitary, Lorentz invariant field theory at low energies due to the constraint $d^2S/dL^2 \leq 0$ \cite{casini2012circle}.

One may ask whether the time-translation symmetry is crucial to obtain the observed transitions. Specifically, consider a circuit where independent unitaries of the form in Eq.\ref{eq:2dim} are applied at each time slice.  In the (unrotated) unitary circuit, as one might expect, breaking time-translational invariance always leads to volume-law entanglement \cite{nahum_2017, khemani_operator_2017, nahum_operator_2018, rakovszky_diffusive_2017,von_keyserlingk_operator_2018, zhou_operator_2018, chen_power_law_2019, Zhou_Nahum_2019}. We confirmed that rotating such a circuit leads to a hybrid projective-unitary circuit that also always exhibits a volume-law scaling. This is because the problem now essentially corresponds to anisotropic bond-percolation in three dimensions where no bonds are removed along one of the directions (namely $y$) and are removed with probability $p$ along the other two directions ($x$ and $t$). Such a model is known to not exhibit a percolation transition for any $p$ \cite{redner1979anisotropic}.

One may also consider the Floquet circuit (Eq.\ref{eq:2dim}) and its space-time dual (Eq.\ref{eq:2dim_rotated}) in 1d. In this case, however, both the unitary circuit and its rotated counterpart are in the area-law phase for any non-zero $p$. To see this, let's consider the unitary circuit and notice that when $p=0$, the spatial support of a single-site Pauli operator grows with time, leading to volume-law entanglement at long times. On the other hand, when $p \neq 0$, there is a finite density of locations (of $O(1/p)$) where the $ZZ$ or the $X$ gates are absent. These locations impose a `wall' such that the end of a stabilizer string cannot grow beyond these walls. This leads to  area-law entanglement $S_A \lesssim O(1/p)$. In contrast, the 2d circuit discussed above allows for a volume-law phase for small non-zero $p$ since a local Pauli stabilizer spreads as a membrane that can bypass the points corresponding to the absent $ZZ$ or $X$ gates. Such a picture suggests that the entanglement transition may be related to a percolation transition, similar to Ref.\cite{Chandran2015semiclassical}. However, the correlation length exponent we numerically obtained differs from the prediction of percolation in two dimensions. It would be be worthwhile to revisit this question in more detail in the future.

\section{Space-time rotation of an interacting Floquet MBL circuit} \label{sec:floquet_mbl}

\begin{figure*}[ht]
	
	\centering
	\begin{subfigure}[b]{0.243\textwidth}
		\includegraphics[width=\textwidth]{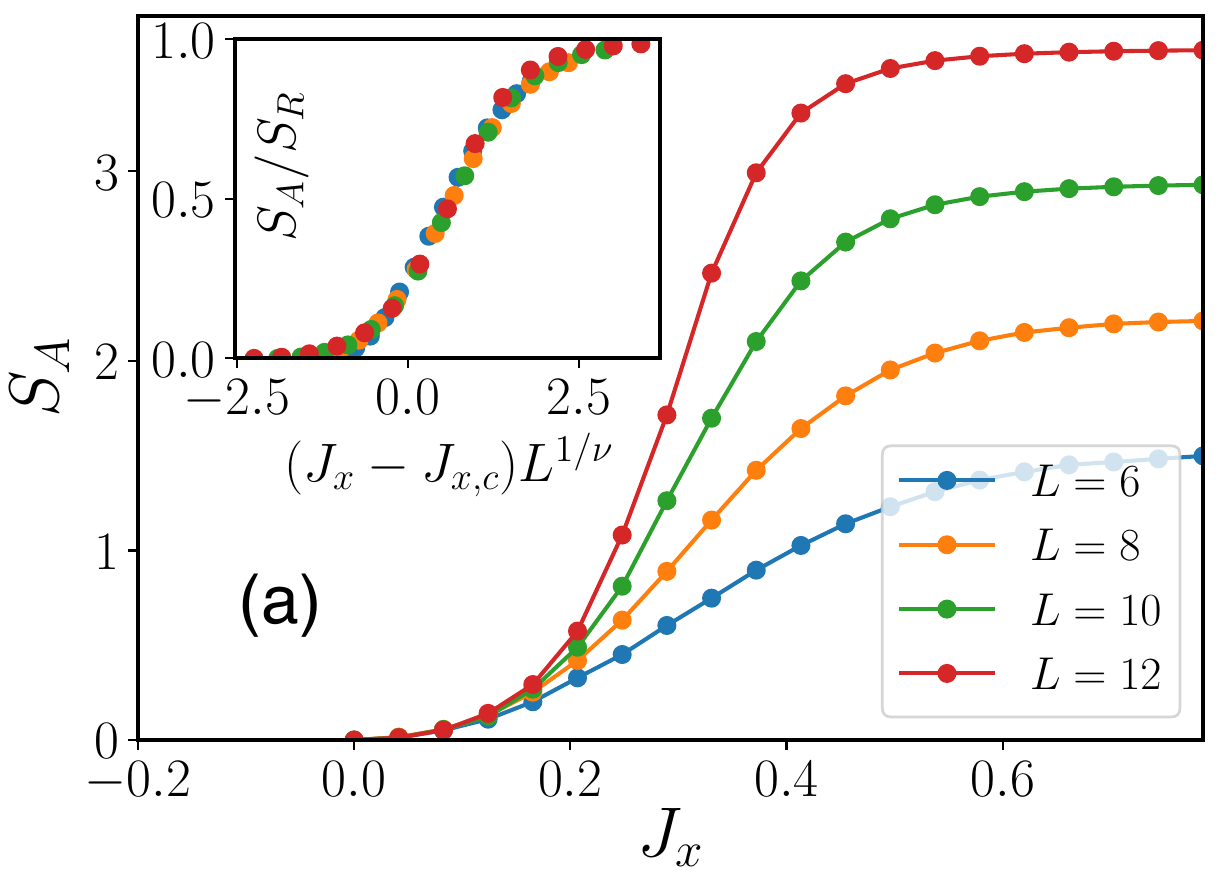}
	\end{subfigure}
	\begin{subfigure}[b]{0.243\textwidth}
		\includegraphics[width=\textwidth]{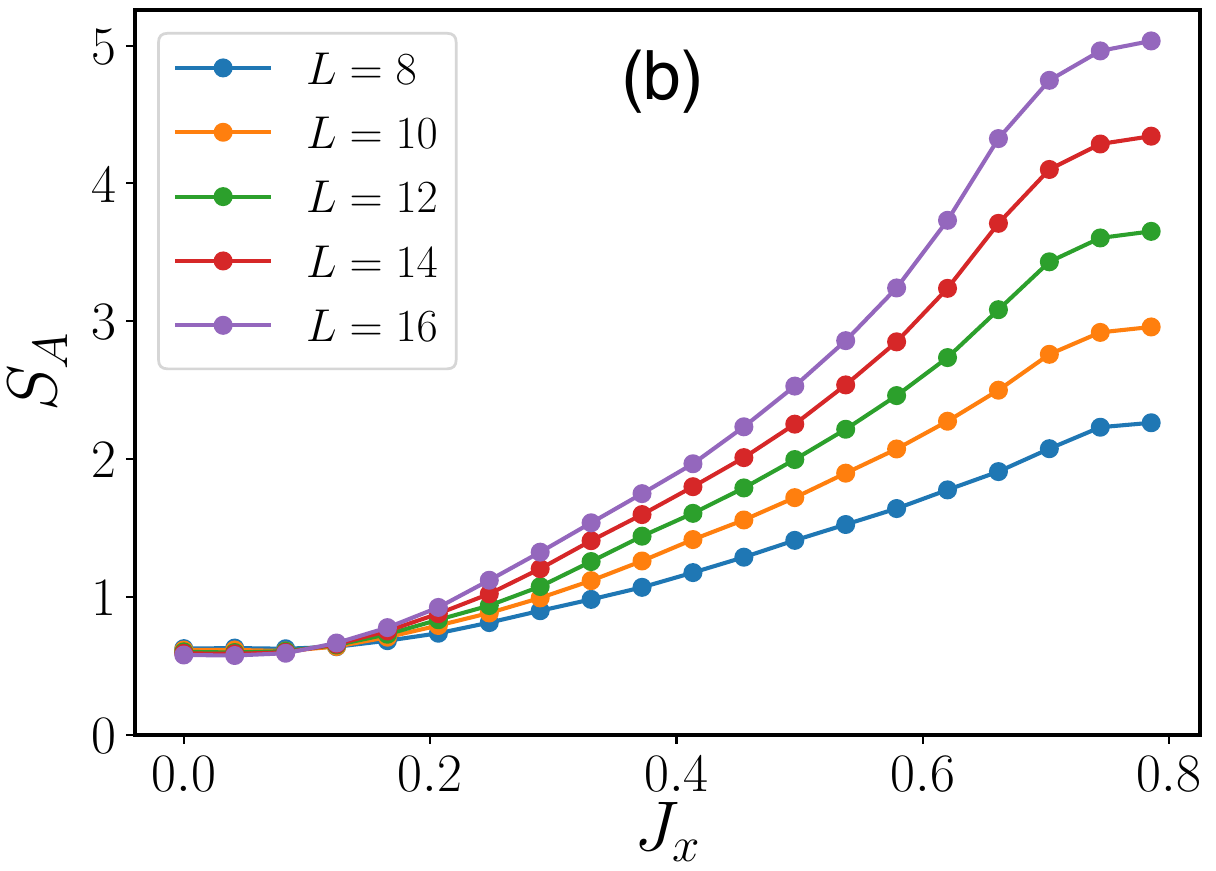}
	\end{subfigure}
	\begin{subfigure}[b]{0.243\textwidth}
		\includegraphics[width=\textwidth]{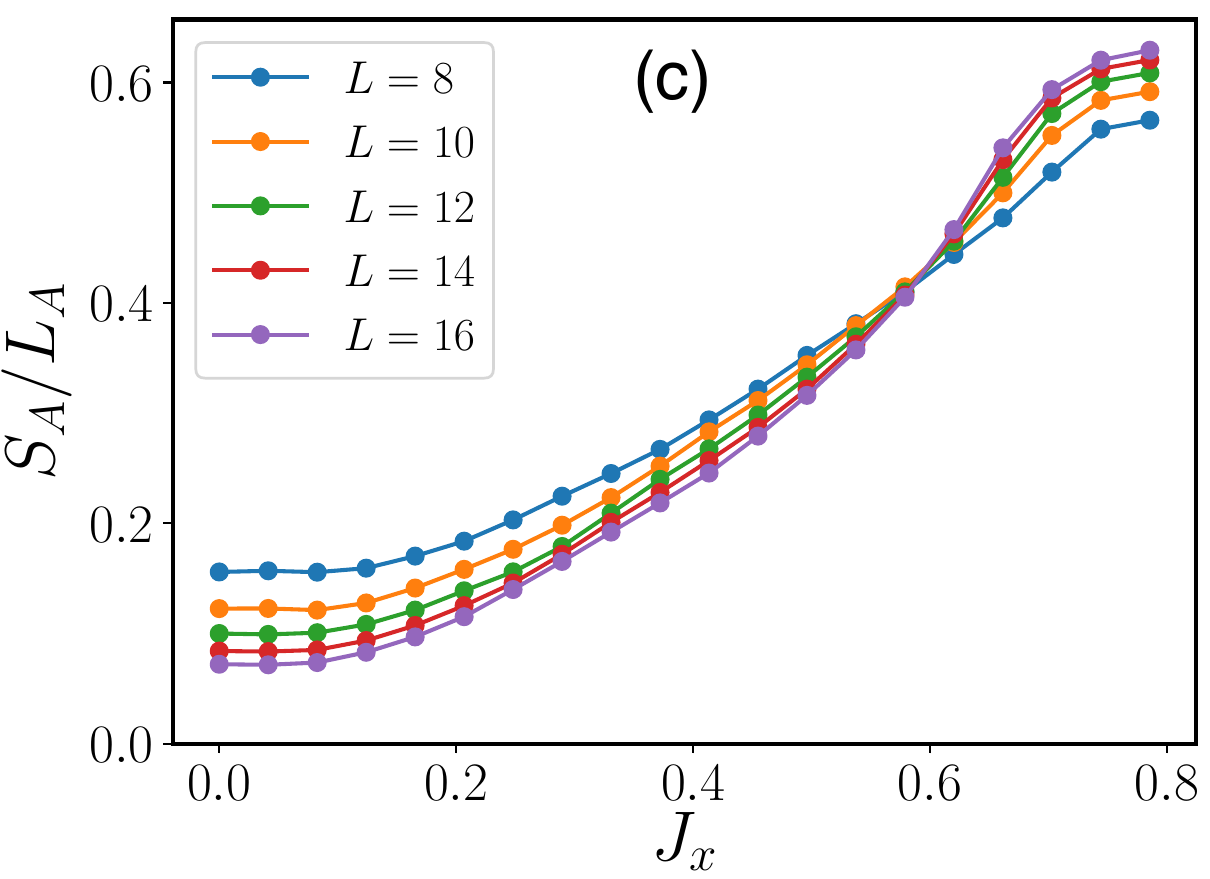}
	\end{subfigure}
	\begin{subfigure}[b]{0.243\textwidth}
		\includegraphics[width=\textwidth]{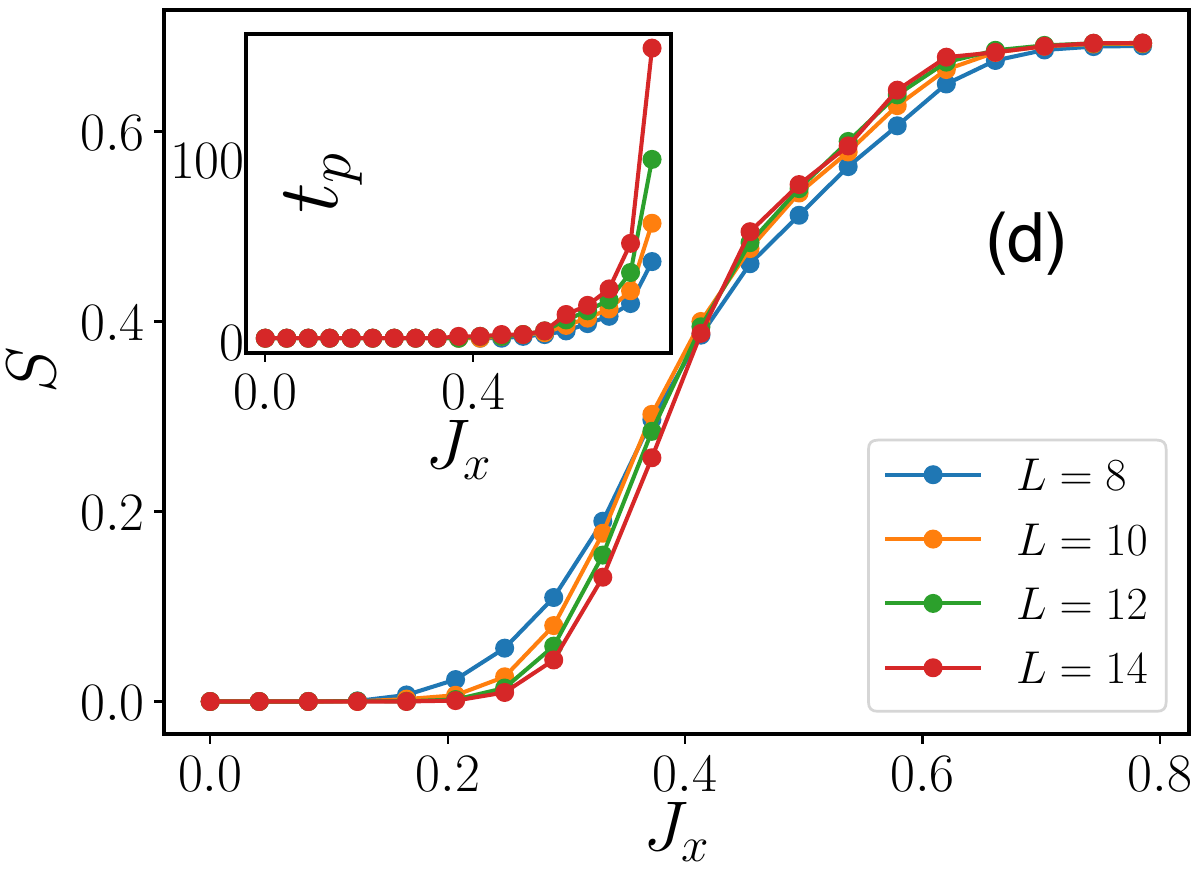}
	\end{subfigure}
	
	\caption{(a) Half-chain entanglement entropy averaged over all eigenstates and 200 random realizations of  the Floquet unitary $U_F$ defined in Eq.\ref{eq:mbl_floquet}. The inset presents the data collapse based on the scaling ansatz: $S_A/S_R= F( (J_x -J_{x,c})L^{1 / \nu})$ with $(J_{x,c},\nu) =( 0.23,1.09  )$ and $S_R= 0.5( L\log2 -1)$ being the entanglement entropy of a random pure state. (b) Entanglement entropy of long-time-evolved states averaged over 150 random realizations of the space-time-rotated non-unitary circuit (Eq.\ref{eq:mbl_rotated_evolve}). (c) Entanglement entropy density of the same circuit as in Fig.(b). (d) The entanglement entropy of an ancilla qubit that is initially prepared in the maximally entangled state with the system, and then evolved for time $t \sim L$ with the non-unitary circuit in Eq.\ref{eq:mbl_rotated_evolve}. Averaging is done over 2000 realizations of the disorder. Inset:  time scale $t_p$ that measures the persistence of the entanglement of the ancilla qubit.}
	\label{fig:interacting_mbl}
\end{figure*}

Finally, we present numerical results on an interacting Floquet model of the form in Eq.\ref{eq:main_floquet}:

\begin{equation}\label{eq:mbl_floquet}
U_F=  e^{ i  J_x \sum_r X_r   }  e^{  - i  \tau \sum_r Z_rZ_{r+1}   - i\tau   \sum_r h_r Z_r }
\end{equation}
where $\tau=0.8$, and $h_r$ is a Gaussian random variable with mean $\overline{h}=0.8090$ and variance $W=1.421$. As shown in Ref.\cite{ Zhang_2016}, tuning $J_x$ induces a transition from an MBL to an ergodic phase, where the Floquet eigenstates exhibit area-law entanglement for small $J_x$ and volume-law entanglement for large $J_x$. Here we study the corresponding space-time dual non-unitary circuit.

As a benchmark, we first confirm the MBL-ergodic transition found in Ref.\cite{Zhang_2016} for the Floquet unitary circuit. Using Exact Diagonalization (ED), we study the half-chain entanglement entropy $S_A$ averaged over all eigenstates of $U_F$, and average the data from 200 random realizations of $U_F$.  We find clear signatures of a transition from a sub-extensive regime to a volume-law regime at finite $J_x=J_{x,c}$. Since eigenstates are localized for $J_x< J_{x,c}$ and are expected to resemble an infinite-temperature pure state (i.e. a random pure/Page state \cite{page1993average} with entanglement entropy $S_R= 0.5( L\log2 -1)$) for any $J_x > J_{x,c}$, we perform a data collapse assuming the scaling form $S_A/S_R= F( (J_x -J_{x,c})L^{1 / \nu})$, and find the critical point $J_{x,c}\approx 0.23$ with the correlation length exponent $\nu=1.09$ (Fig.\ref{fig:interacting_mbl} (a) inset).

The space-time-rotated circuit corresponding to $U_F$ is
\begin{equation}\label{eq:mbl_rotated_evolve}
V(T) = \prod_{t=1}^T V_t, \quad V_t=  e^{  i  \tilde{J}_x \sum_r X_r  } e^{ i   \tilde{J}_z  \sum_r Z_rZ_{r+1}   -i  \tau h(t)\sum_r  Z_r },
\end{equation} 
where the field $h$ is now random in the time direction due to the space-time rotation, and the couplings $\tilde{J}_x, \tilde{J}_z$ are defined in Sec.\ref{sec:spacetime}. We first analyze the entanglement structure of states evolved via $V$ for times $T \sim L $.  We find signatures of a transition by tuning $J_x$ (see Fig.\ref{fig:interacting_mbl} (b)). In particular, when one plots entanglement entropy \textit{density}, one finds a crossing at $J_x \approx 0.6$ (see Fig.\ref{fig:interacting_mbl} (c)), which separates a regime with volume-law entanglement from a regime where the entanglement is sub-extensive.  

Finally, we  study the entanglement dynamics of an ancilla qubit that is initially maximally entangled with the system, following the protocol in Refs.\cite{gullans2019dynamical,gullans2019scalable}. We evolve the system for time $T \sim L$, and find a crossing around $J_x \approx 0.4$ (see Fig.\ref{fig:interacting_mbl} (d)). In addition, the entanglement $S$ of the ancilla qubit shows distinct features on two sides of this crossing (see Appendix.\ref{appendix:mbl} for numerical data). For $J_x\lesssim 0.4$, the entanglement entropy of the ancilla qubit decays from its initial value ( $=\log 2$) exponentially with time, while for $J_x \gtrsim 0.4$, it remains at its initial value for a while (i.e. exhibits a `plateau'), followed by an exponential decay. To quantify the plateau interval, we define a `purification time' $t_p$ as the time after which the entanglement of the ancilla qubit has dropped below $0.65$ ($\approx 0.94 \log(2)$). We find $t_p\approx O(1)$ for $J_x \lesssim 0.4$ while it increases with system size $L$ for $J_x \gtrsim 0.4$ (Fig.\ref{fig:interacting_mbl} inset). Notably, for large enough $J_x (\gtrsim 0.7)$, we find that $t_p$ grows super-linearly with $L$, and therefore the non-unitary circuit may potentially serve as a good quantum error-correcting code.

Finally, we note that different values of the crossing points in different measures suggest that the finite size effects are likely strong at these system sizes. However, at the very least, the trends strongly indicate a stable volume law phase at $J_x \gtrsim 0.6$ (see Fig.\ref{fig:interacting_mbl} (c)), and a phase with sub-extensive entanglement at small but non-zero $J_x$.

\section{Space-time rotated correlators: post-selection free measurement and physical consequences } \label{sec:correlations}

\begin{figure}[h]
	\centering
	\includegraphics[width=0.55\hsize]{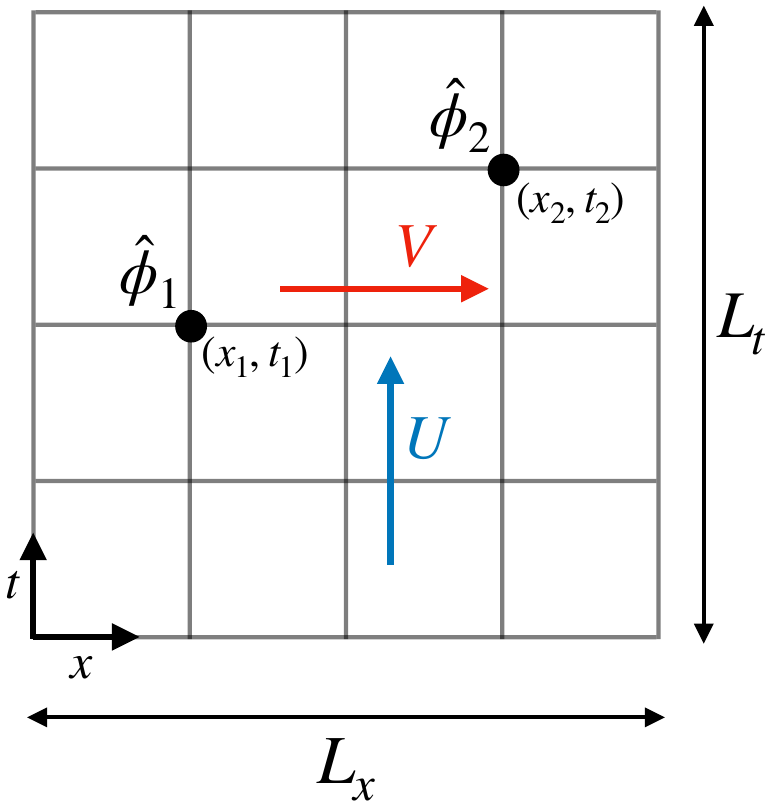}
	\caption{Geometry for the correlation function $C(x_1,t_1;x_2,t_2)$ defined in the main text (Eq.\ref{eq:correlator_action}). It can be interpreted in two different ways: either as a correlation function for a system evolving unitarily with circuit $U$, or as a correlation function for a system evolving with the rotated non-unitary circuit $V$, see Eq.\ref{eq:correlator_operator}.}
	\label{fig:operator_insert}
\end{figure}

Since the circuits related by space-time rotation have the same bulk action $S$ (see the Introduction and Fig.\ref{fig:circuitscheme}), it is natural to seek a relation between their physical observables. At the outset, one notices that conventional correlation functions such as $\langle \psi_0| U^{\dagger} O U |\psi_0\rangle$ in the unitary circuit are \textit{not} related to similarly defined correlations functions in its space-time rotated non-unitary $V$, such as $\langle \psi'_0| V^{\dagger} O V |\psi'_0\rangle/ \langle \psi'_0| V^{\dagger} V |\psi'_0\rangle$. Referring to Fig.\ref{fig:circuitscheme}, this is because in the former case,  the fields $\phi_0$ and $\phi_t$ are held fixed to define the wavefunction, and the fields $\phi'_0$ and $\phi_x$ are being summed over, while in the latter case, it is the other way around. However, consider the following object (see Fig.\ref{fig:operator_insert}):

\be 
C(x_1,t_1;x_2,t_2) = \frac{\int D\phi \,\,\phi_1(x_1,t_1) \phi_2(x_2,t_2) e^{i S(\{\phi\})}}{\int D\phi \,\,e^{i S(\{\phi\})}} \label{eq:correlator_action}
\ee 
Since the action $S$ is invariant under space-time rotation and one is summing over all fields in the above integral, $C$ has a well-defined meaning in both the rotated and unrotated circuits:

\beq
C(x_1,t_1;x_2,t_2) & = & \frac{\tr \left( U(t_2,t_1) \hat{\phi}_1 U(t_1,t_2) \hat{\phi}_2 \right)}{\tr \left( U(t_2,t_1) U(t_1,t_2) \right)} \nonumber \\
& = &  \frac{\tr  \left(  V(x_2,x_1) \hat{\phi}_1  V(x_1,x_2)  \hat{\phi}_2 \right)}{\tr \left(V(x_2,x_1) V(x_1,x_2)\right) }  \label{eq:correlator_operator}
\eeq 
where $U(a,b)$ and $V(a,b)$ are evolution operators from time $a$ to $b$ when $ a < b$, while when $a > b$,  $U(a, b)  = U(a,L_t)U(0,b)$, and $V(a, b) = V(a, L_x)V(0,b)$ (see Fig.\ref{fig:operator_insert} for definitions of $L_t, L_x$). $\hat{\phi}_1, \hat{\phi}_2$ are operators corresponding to the fields $\phi_1, \phi_2$ in Eq.\ref{eq:correlator_action}, whose space-time insertion locations are shown in Fig.\ref{fig:operator_insert}.

The correlation functions in Eqs.\ref{eq:correlator_action}, \ref{eq:correlator_operator} are rather unconventional since there is no `backward trajectory' as in the standard Keldysh expression \cite{keldysh1965diagram, kadanoff1962quantum} for conventional correlation functions such as $\langle \psi_0| U^{\dagger} O U |\psi_0\rangle$. To measure such correlators experimentally, one may employ the idea of a control qubit that generates two branches of  a many-body state \cite{jiang2008anyonic, abanin2012measuring, muller2009mesoscopic}. For example, to measure $\langle \psi_0| U_1 \hat{\phi}_1 U_2 \hat{\phi}_2 |\psi_0\rangle$ for some $U_1, U_2$ and a product state $|\psi_0\rangle$, the total system is initially prepared in a state $|\psi_0\rangle \otimes \left(|\uparrow \rangle + |\downarrow\rangle \right)$ where the expression after $\otimes$ denotes the state of the control qubit. Using standard techniques \cite{jiang2008anyonic, abanin2012measuring, muller2009mesoscopic}, one then applies the operator $U_2 \hat{\phi}_2$ on the `up-branch' of this initial state, i.e., the state $|\psi_0\rangle \otimes |\uparrow \rangle$, and similarly, one applies the operator $ \hat{\phi}^{\dagger}_1 U^{\dagger}_1$ on the down branch. Finally, one measures, the expectation value of the $\sigma^x$ and the $\sigma^y$ operators that act on the control qubit, which yields the object of interest, namely, the real and imaginary parts of $\langle \psi_0| U_1 \hat{\phi}_1 U_2 \hat{\phi}_2 |\psi_0\rangle$. The trace in Eq.\ref{eq:correlator_operator} would then need to be approximated by sampling over several such expressions, although even a single/few such expressions may sometime capture the qualitative aspects of interest (see below).

\begin{figure}[h]
	\centering
	\includegraphics[width=0.8\hsize]{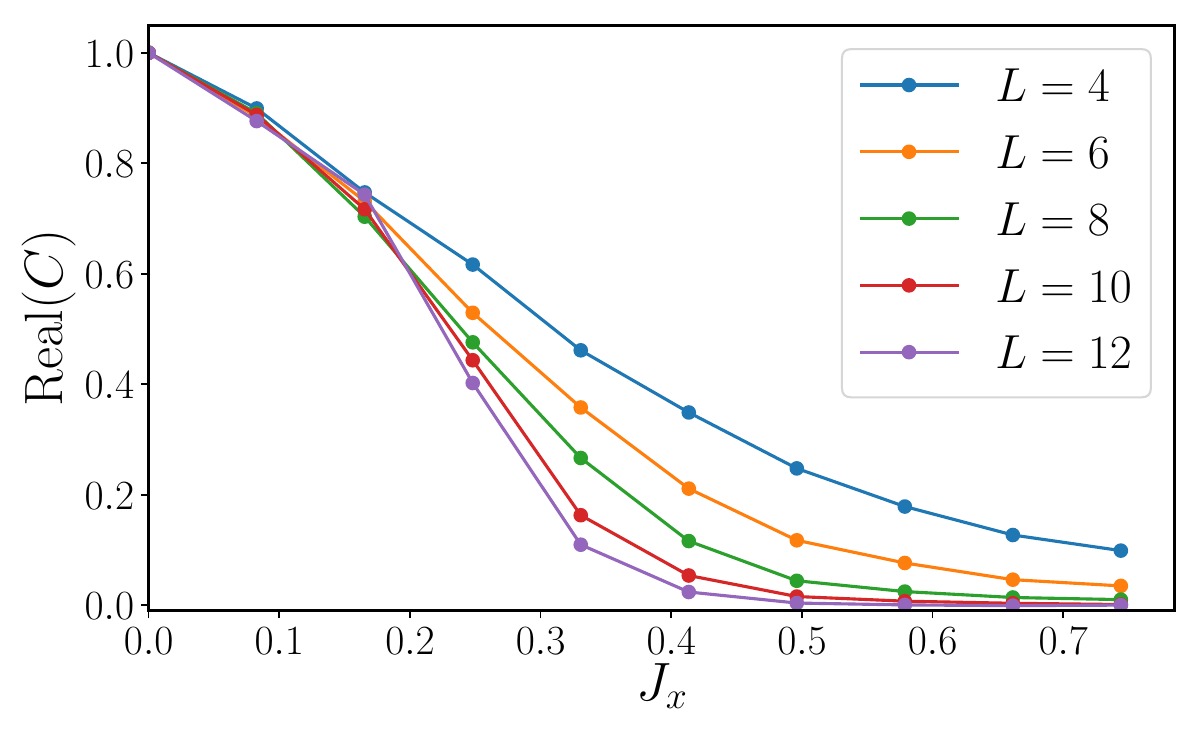}
	\caption{Correlation function $C(r,0;r,T/2)$ for $T \gtrsim O(L)$ in the circuit $U_F$ defined in Eq.\ref{eq:mbl_floquet}. The  data presented is obtained by averaging over 2560 random realizations of $U_F$ for $L=4,6,8,10$, and 512 random realizations of $U_F$ for $L=12$.   }
	\label{fig:unrotated_correlator_Jx}
\end{figure}

Due to the unconventional nature of the correlator $C$, it is not obvious if it captures universal physics. We now provide a heuristic argument that $C$ is singular across the MBL transition. We recall that an MBL system hosts emergent `$\ell$-bit' degrees of freedom \cite{huse_lbits,abanin_lbits} $\{\tau^z\}$ which have a non-zero overlap with the local $\sigma^z$ operators: $\sigma_i^{z} = Z \tau^{z}_i + ...$, where $Z \neq 0$ denotes the overlap between $\sigma_i^{z}$ and $\tau^{z}_i$ (`quasiparticle residue'). In particular, this implies that at long times $\sigma_i^{z}(t) = Z \sigma_i^{z}(0) + ...$, i.e., $\sigma_i^{z}(t)$ has a non-zero overlap with $\sigma_i^{z}(0)$.  Let us use this fact to simplify the numerator of $C$ in Eq.\ref{eq:correlator_operator}: $\tr \left( U(0,T/2) \sigma^z_r U(T/2,0) \sigma^z_r \right)  = \tr \left( \sigma^z_r(T/2) U(0,T/2)  U(T/2,0) \sigma^z_r\right) = Z \tr \left( U(0,T/2)  U(T/2,0)\right) + ...$ where `...' denotes terms that are expected to vanish at long times after averaging over time and disorder. This suggests that in an MBL phase, $C(r,0;r,T/2) $ simply equals $Z$ at large $T$, and therefore vanishes as one approaches the delocalization transition. We verified this expectation numerically using ED for the Floquet model studied in Sec.\ref{sec:floquet_mbl}, see Fig.\ref{fig:unrotated_correlator_Jx}. We also calculated a simpler correlator, namely, $\langle \psi_0| U(0,T/2) \sigma^z_r U(T/2,0) \sigma^z_r |\psi_0\rangle$, where $|\psi_0\rangle$ is a product state, and found that it behaves quite similarly to $C$.

We also studied the correlator in the 1d free-fermion circuit discussed in Sec.\ref{sec:1d_floquet_ising} as well as the 2d Clifford circuit discussed in Sec.\ref{sec:2d_floquet_clifford}. We found that the correlator fails to distinguish between the localized phase and the delocalized phase in either of these circuits for distinct reasons. For a localized free-fermion circuit, the time-evolved operator $\sigma_i^z(t)$ continues to have a non-zero overlap $Z$ with $\sigma_i^z$ at arbitrarily long times, i.e. $\sigma_i^z(t) = Z \sigma_i +...$.  However, due to the lack of dephasing in free-fermion circuits (see e.g. Ref.\cite{Moore_2015_mbl}), the terms under `$...$' do not vanish even at long times despite averaging over disorder, and their contribution fluctuates in time significantly at all times. Consequently, the spacetime-rotated correlator does not provide a clear signature across the localization transition. 

On the other hand, for a Clifford circuit, the correlator $C$ does not differentiate between a localized phase and a delocalized phase due to the absence of the `$\ell$-bit' picture $\sigma_i^z(t)=Z \sigma_i^z + ...$ . Specifically, $\sigma_i^z(t)$ will always be a \textit{single} product of Pauli operators over various sites, and the localization/delocalization phase manifests in the bounded/unbounded spatial support of $\sigma_i^z(t)$, instead of the relative weight of various operators. Therefore, our aforementioned argument in the context of generic MBL systems does not apply.

We note that Ref.\cite{Ippoliti_2021b} discussed an alternative method to relate quantities between a unitary circuit and its rotated non-unitary counterpart. In particular, Ref.\cite{Ippoliti_2021b}  considered a protocol where the purification dynamics in the non-unitary circuit can be obtained by a combination of unitary dynamics and projective measurements.

\section{Summary and Discussion} \label{sec:discuss}

In this work, we employed the idea of the space-time rotation of unitary circuits to construct non-unitary circuits that display entanglement phase transitions. We  focused on specific Floquet unitary circuits that display localization-delocalization transitions of various kinds (free fermion, Clifford, many-body). We  found that the delocalized (localized) regime of the unitary circuit maps to a regime with volume-law (area-law/critical) entanglement in the corresponding non-unitary circuit. Therefore, the space-time rotation maps the physics of localization to the physics of quantum Zeno effect. We also found that the entanglement transitions in the non-unitary circuits are accompanied by purification transitions of the kind introduced in Refs.\cite{gullans2019dynamical,gullans2019scalable}. We  introduced an unconventional correlator in the non-unitary theory that can in principle be measured without requiring any post-selection, and provided a  heuristic argument that this correlator is singular across an MBL transition. 

Our procedure leads to the construction of a non-unitary free fermion circuit that supports volume-law entanglement, which has hitherto been elusive \cite{Chen_2020, tang2021quantum, jian2020criticality, jian2021yanglee}. As discussed in Sec.\ref{sec:1d_floquet_ising}, we find that a non-unitary circuit obtained by the rotation of a free fermion unitary circuit has the special property that the real parts of certain hopping elements are automatically pinned to $\pi/4$. This leads to volume-law entanglement when the non-unitary circuit has translational symmetry in both space and time, and the possibility of a volume-law to area-law transition when disorder is introduced in the non-unitary circuit along the time direction.

Given our results, it is natural to ask if the space-time rotation of a unitary circuit $U$ hosting a localization-delocalization transition always leads to a non-unitary circuit $V$ that also shows an entanglement transition. Firstly, we note that a localization-delocalization transition in a unitary system will induce a singularity in the spectral form factor since the spectral form factor is well known to be sensitive to quantum chaos. Due to our mapping, the spectral form factor for the non-unitary theory will also be singular across the transition (since $= |\tr U|^2 = |\tr V|^2$). Recent progress \cite{li2021spectral} shows that at least for a class of non-unitary evolution, the spectral form factor continues to encode features of quantum chaos.  Further, as discussed in Sec.\ref{sec:correlations}, a correlator that is well-defined in both the unitary and the non-unitary theory can be argued to be singular across an MBL transition. However, this correlator is a bit hard to interpret physically within the non-unitary theory. It will be worthwhile to pursue a physical understanding of the spectral form factor and the correlator in Sec.\ref{sec:correlations} for the non-unitary theories studied in this paper.

We also explored the role played by the time-translation symmetry of the unitary circuit. In the examples we studied, breaking of time-translation symmetry also leads to the absence of entanglement transition in the rotated non-unitary circuit. We suspect that the entanglement transitions in  non-unitary circuits that are space-time dual of  time-translationally invariant unitary circuits belong to a different universality class compared to those hosted by non-unitary circuits where such a symmetry is absent. 

As argued in Ref.\cite{Ippoliti_2021b}, if a non-unitary circuit is related to a unitary circuit via space-time rotation, then at least some of its properties (such as the purification rate) may be obtained purely via unitary evolution combined with a small number of projective measurements. Furthermore, as discussed in Sec.\ref{sec:correlations}, an unconventional correlator in the non-unitary theory can be measured using only unitary operations. Applying these results to  the examples discussed in this work would potentially allow one to access the physics of entanglement transitions in hybrid projective-unitary circuits without post-selection. 

We note that Ref.\cite{jian2020criticality} introduced an interesting relation between non-unitary circuits of free fermions in $d+1$ space-time dimensions and the Anderson localization-delocalization transition for Hermitian Hamiltonians in $d+1$ \textit{space} dimensions. The basic idea employed is to relate the circuit in $d+1$ space-time dimensions to the scattering matrix that describes the Chalker-Coddington model \cite{chalker1988percolation} in $d+1$ dimensional space. In contrast, our work focuses on relating a unitary and a non-unitary system that live in the same number of space-time dimensions. It might be worthwhile to understand the volume-law phase in our non-unitary circuit of free fermions (Sec.\ref{sec:1d_floquet_ising}) and its higher dimensional generalizations from the perspective in Ref.\cite{jian2020criticality}.

\emph{Note Added:} After the completion of this work, we became aware of a related work ~\cite{Ippoliti2021fractal} (appearing in the same arXiv posting) which also considers entanglement dynamics in spacetime duals of unitary circuits. Our works are largely complementary and agree where they overlap.

\acknowledgements{
	We are grateful to John McGreevy and Yahya Alavirad for illuminating discussions and helpful comments on the draft. We  thank Matteo Ippoliti, Tibor Rakovszky and Vedika Khemani for pointing out an incorrect statement in the Section `Summary and Discussion' of the first version of this pre-print, see footnote  \cite{footnote:fisherpaper} for details. TG acknowledges support by the National Science Foundation under Grant No. DMR-1752417, and by an Alfred P. Sloan Research Fellowship. This work used the Extreme Science and Engineering Discovery Environment (XSEDE)~\cite{xsede}, which is supported by National Science Foundation grant number ACI-1548562. We acknowledge support from the University of California’s Multicampus Research Programs and	Initiatives (MRP-19-601445).
}

\bibliography{v1bib}
\renewcommand\refname{Reference}
\bibliographystyle{unsrt}
\newpage

\appendix

\onecolumngrid

\section{Additional details on 1+1D Floquet quasiperiodic circuit}\label{appendix:AA_ising}

\subsection{Entanglement entropy}\label{appendix:AA_ising_entanglement}
Here we outline the calculation of entanglement entropy of time-evolved states in the 1+1D circuit (Eq.\ref{eq:AA_floquet}): 
\begin{equation}\label{eq:appendix_AA_floquet}
U_F =e^{ i J \sum_j  X_jX_{j+1}  }e^{ i  \sum_j h_jZ_j},
\end{equation}
where $h_j= h+ \lambda  \cos( 2\pi Q j + \delta)$, and $Q=\frac{2}{1+\sqrt{5}}$. 

We first map the circuit to a fermionic model using the Jordan-Wigner transformation:
\begin{equation}
Z_i= 1-2c_i^{\dagger} c_i, \quad c_i^{+} = \left( \prod_{j=1}^{i-1} Z_j   \right) \sigma_i^-  , \quad c_i^{-} = \left( \prod_{j=1}^{i-1} Z_j   \right) \sigma_i^+.
\end{equation}
Correspondingly, $X_i= \sigma_i^- +\sigma_i^{+}  = \left( \prod_{ j=1  }^{i-1} (1- 2c_j^{\dagger} c_j  )    \right) ( c_i +c_i^{+} ) $, and 

\begin{equation}
\sum_{i=1}^L X_iX_{i+1}  = \sum_{i=1}^{L-1} (c_i^{\dagger} -c_i )(c_{i+1}^{\dagger} + c_{i+1} )  -  e^{i\pi N } (  c^{\dagger}_L-  c_L   )  (  c^{\dagger}_{1}+  c_{1}  )  =  \sum_{i=1}^L  (c_i^{\dagger} -c_i )(c_{i+1}^{\dagger} + c_{i+1} ), 
\end{equation}
where $e^{i \pi N}$ measures the total fermion number parity: $e^{i \pi N} = e^{i\pi  \sum_i c_i^{\dagger}c_i   } = \prod_{i=1}^L (1-2c^{\dagger}_ic_i)$. We impose antiperiodic boundary condition, $c_{L+1}=-c_1$, for even fermion parity sector and periodic boundary condition, $c_{L+1}=c_1$, for odd fermion parity sector.

Since the Floquet dynamics does not conserve the total fermion number, it is more convenient to employ the Majorana fermions by defining $a_{2j-1 } = c_j + c_j^{\dagger}  $ and $a_{2j } = i (c_j -  c_j^{\dagger} )$, which satisfy $\{a_i,a_j   \} = 2\delta_{ij}$. The Floquet unitary defined in Eq.\ref{eq:appendix_AA_floquet} then reads

\begin{equation}
U_F= U_{XX} U_Z= e^{-J \sum_{j=1}^L a_{2j}a_{2j+1} }e^{ \sum_{j=1}^L h_j a_{2j-1}a_{2j} }.
\end{equation}

Since $U_F$ is Gaussian in Majorana fermions, the Majoranas evolve under $U_F$ as

\begin{equation}
U_F^{\dagger} a_i  U_F= U_Z^{\dagger} U_{XX}^{\dagger} a_i U_{XX}U_Z  = \sum_k  O_{ik }a_k,
\end{equation}
where $O$ is an orthogonal matrix. Correspondingly, the Majoranas at time $t$ can be obtained by repeatedly applying the orthogonal transformation on $\{a_i\}$: $a_i(t) = \sum_j (O^t)_{ij}a_j$. Using this formalism, we can calculate the correlation matrix at time $t$: $\Gamma_{ij}(t) =  \expval{a_i(t) a_j(t)   }  -\delta_{ij}$, from which the entanglement entropy between a region $A$ and its compliment can be found by diagonalizing $\Gamma_A(t)$, the restriction of the correlation matrix to the region $A$ (Refs. \cite{Chung2001density, cheong2004manybody, Peschel_2003}): 

\begin{equation}
S_A= - \sum_{i=1}^{L_A} \left[ \frac{1-\nu_i}{2 } \log\left(   \frac{1-\nu_i}{2 }  \right)  +    \frac{1+\nu_i}{2 } \log\left(   \frac{1+\nu_i}{2 }  \right)   \right].  
\end{equation}
with  $\{\pm \nu_i\} $ being the $2L_A$ eigenvalues of $\Gamma_A(t)$.

\subsection{Single-particle eigenfunctions of Floquet unitary}\label{appendix:AA_single_particle}
\begin{figure}[t]
	\centering
	\begin{subfigure}[b]{0.42\textwidth}
		\includegraphics[width=\textwidth]{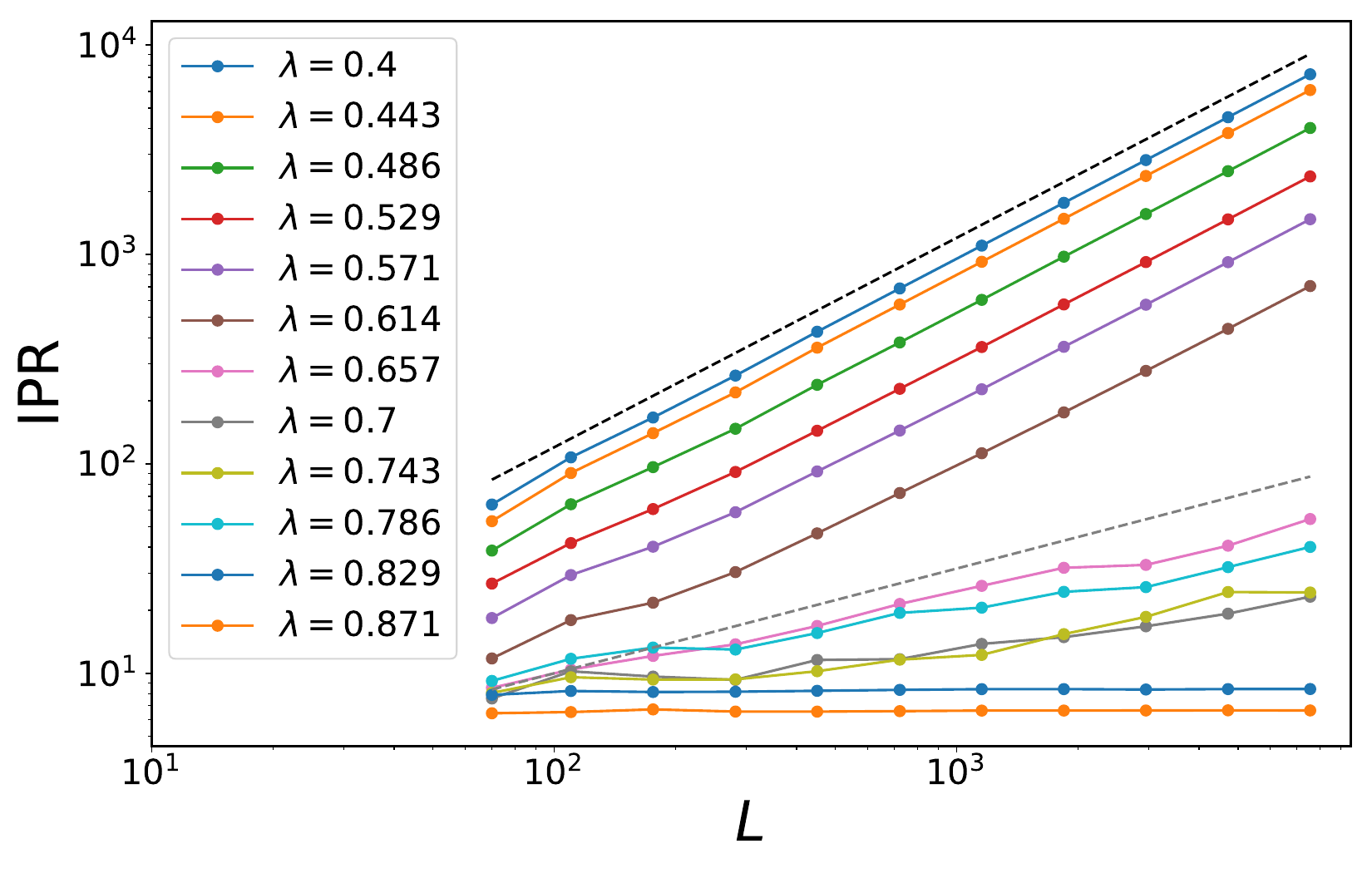}
	\end{subfigure}
	\begin{subfigure}[b]{0.42\textwidth}
		\includegraphics[width=\textwidth]{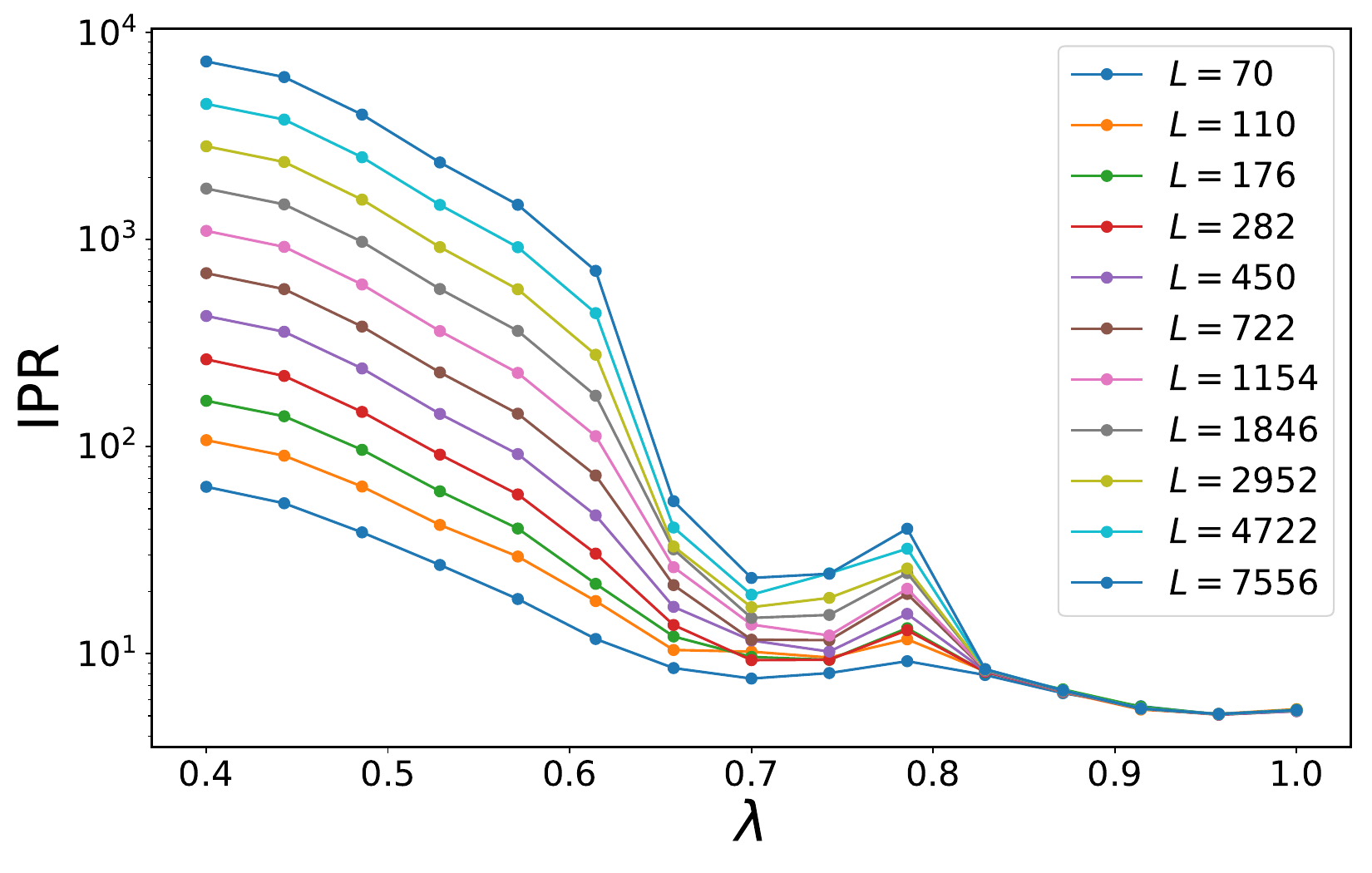}
	\end{subfigure}
	\caption{Inverse participation ratio IPR averaged over all eigenstates of the orthogonal matrix $O$ which governs the dynamics of Majorana fermions up to the total system size $L=7556$. In the left panel, the black and gray dashed lines serve as a reference for the scaling laws $\text{IPR}\sim L$ and $\text{IPR} \sim  \sqrt{L}$ respectively. }  \label{fig:un_ipr}
\end{figure}

Here we discuss the properties of the single-particle eigenfunctions of $U_F$ in terms of the Majorana fermions. We find signatures of three distinct phases, in line with the results from entanglement entropy of long-time-evolved states (Sec.\ref{sec:1d_floquet_ising} in the main text). Specifically, we study the inverse participation ratio: 
\begin{equation}
\text{IPR}=\frac{1}{ \sum_i \abs{\psi_i}^4   },  
\end{equation}
where $\psi_i$ is an eigenfunction of $O$ at the $i$-th Majorana site. We recall that the IPR is a conventional tool to quantify the localization/delocalization property of wavefunctions. In one spatial dimension, an extended (delocalized) wavefunction has $\abs{\psi_i} \sim O(1/\sqrt{L})$, implying $\text{IPR} \sim O(L)$. On the other hand, a localized wavefunction is mainly supported on a finite number of lattice sites, yielding $\text{IPR} \sim O(1)$. Here we study the IPR averaged over all eigenstates of $O$, and find that the averaged IPR (denoted as $\expval{\text{IPR}}$) exhibits three different scalings with the system size $L$ as the modulation strength $\lambda$ is varied, similar to the entanglement entropy of long-time-evolved many-body states. For small $\lambda$, $\expval{\text{IPR}}$ scales as $O(L)$, a signature of a delocalized phase, while for large $\lambda$, $\expval{\text{IPR}}\sim O(1)$, corresponding to a localized phase. In addition, there is an intermediate regime ($0.64 \lesssim \lambda \lesssim  0.8$), where $\expval{\text{IPR}}$ scales as $O(L^{\gamma})$ with $\gamma \sim 0.5$ (see Fig.\ref{fig:un_ipr}).

\subsection{Scaling collapse of entanglement entropy}\label{appendix:collapse}
\begin{figure}[h]
	\centering
	\begin{subfigure}[b]{0.35\textwidth}
		\includegraphics[width=\textwidth]{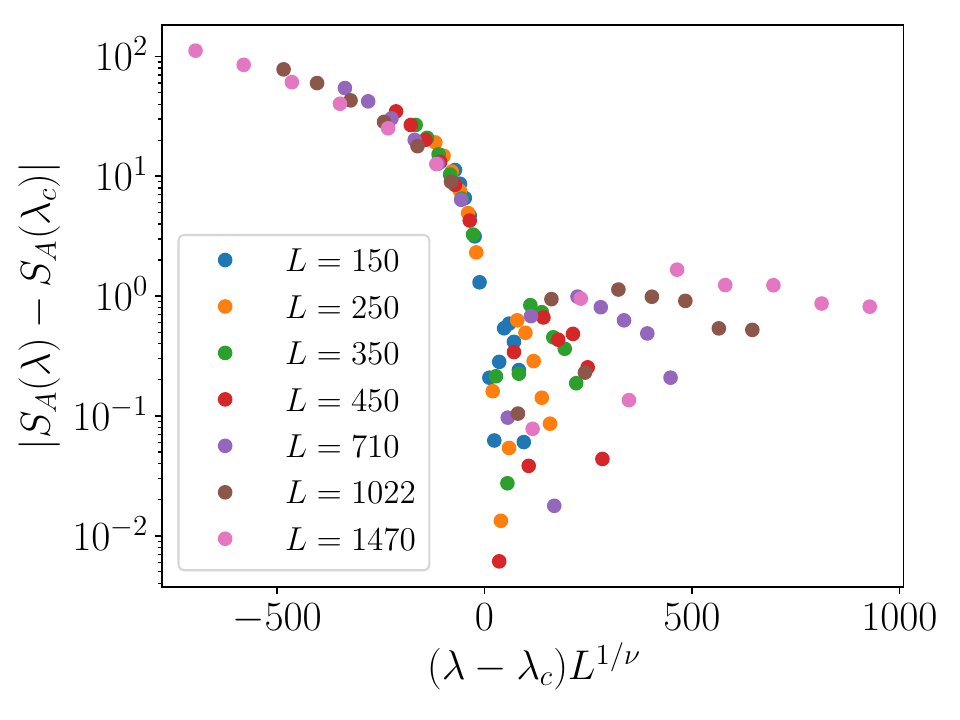}
	\end{subfigure}
	\caption{Scaling collapse of the late-time entanglement entropy for the rotated circuit defined in Eq.\ref{eq:1d_rotated_circuit}. We use the scaling ansatz $\abs{S_A(\lambda) -S_A( \lambda_c ) }  =F( (\lambda-  \lambda_c )L^{1/\nu} )$  where $(\lambda_{c}, \nu) \approx (0.64, 1.0)$.  
	}  \label{fig:appendix_collapse}
\end{figure}

In Fig.\ref{fig:appendix_collapse} we provide numerical data for the scaling collapse of the late-time entanglement entropy for the rotated circuit (Eq.\ref{eq:1d_rotated_circuit}).

\subsection{Purification dynamics}\label{appendix:AA_purification}
\begin{figure}[h]
	\centering
	\begin{subfigure}[b]{0.33\textwidth}
		\includegraphics[width=\textwidth]{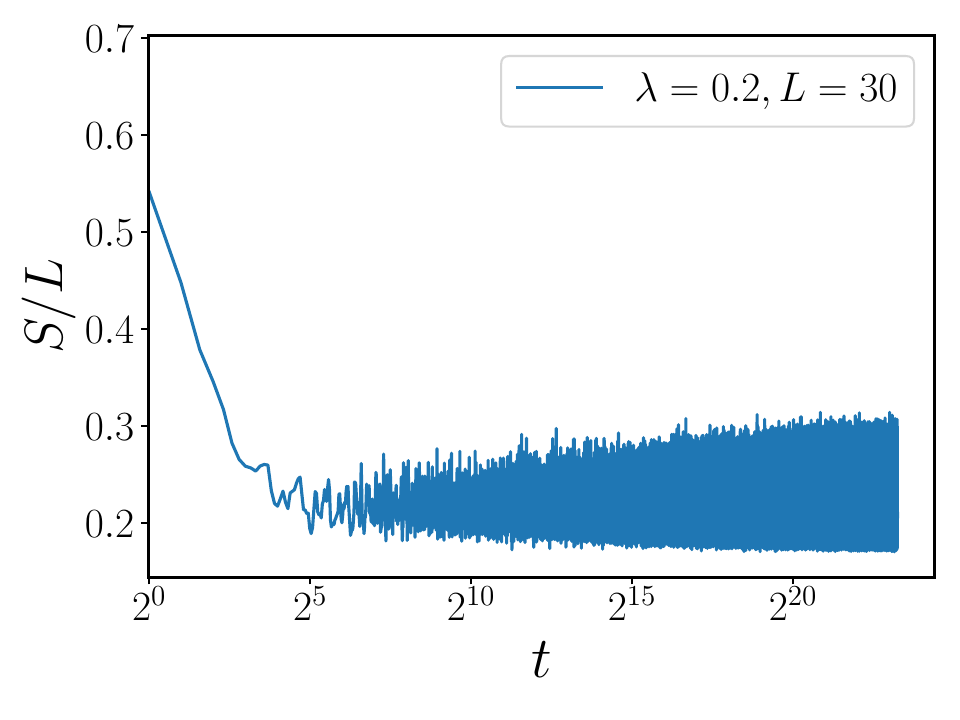}
	\end{subfigure}
	\begin{subfigure}[b]{0.33\textwidth}
		\includegraphics[width=\textwidth]{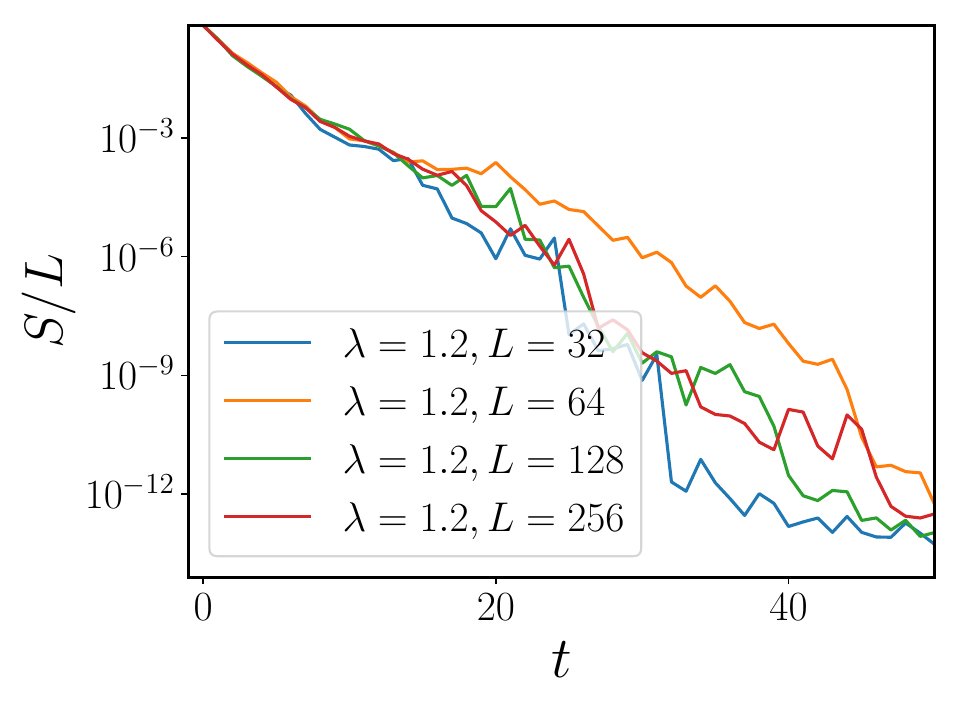}
	\end{subfigure}
	\caption{Time evolution of entropy density for the non-unitary circuit defined in Eq.\ref{eq:1d_rotated_circuit}.	The density matrix is  prepared at $t = 0$ in a completely mixed state, $\rho(t=0) \propto \mathbb{1} $.
	}  \label{fig:appendix_purification_dynamics}
\end{figure}

Here we present additional numerical results on the purification dynamics of a density matrix that is initially in a completely mixed state (i.e. $\rho(t=0) \propto \mathbb{1} $) and is evolved with the non-unitary circuit defined in Eq.\ref{eq:1d_rotated_circuit} (see Fig.\ref{fig:appendix_purification_dynamics}). At $\lambda=0.2$ (i.e. in the volume-law phase), entropy density $S/L$ decreases at short times and remains non-zero for the longest observed time ($t \sim 2^{L})$. At $\lambda=1.2$ (i.e. in the critical phase), entropy density decreases exponentially to zero within a characteristic time scale that is independent of $L$.

\section{1+1D non-unitary circuits with spacetime translational invariance}\label{appendix:pi_over_four}
Given the Floquet unitary 
\begin{equation}
	U_F= e^{  i J\sum_jX_jX_{j+1}} e^{   i h \sum_j Z_j},
\end{equation}
we show that when the real part of both $J$ and $h$ is $\pi/4$, a time-evolved state at a typical long time exhibits volume-law entanglement. We solve the model using the standard Jordan-Wigner transformation, where we first introduce the complex fermions $c_j$ in real space, and then Fourier transform the fermions to momentum space $c_k=\frac{1}{\sqrt{L}} \sum_j  e^{-ikj }c_j$. It follows that $U_F$ can be written as 
\begin{equation}
	U_F  =\prod_{k>0} \exp{ 2iJ  \begin{pmatrix} c_k^{\dagger}  &c_{-k}   \end{pmatrix}   \begin{pmatrix}  \cos k &  i \sin k \\  -i  \sin k &  - \cos k     \end{pmatrix}   \begin{pmatrix}   c_k\\ c_{-k}^{\dagger}     \end{pmatrix} }  \exp{ -2ih \left(  c_k^{\dagger}c_k+ c_{-k}^{\dagger}c_{-k}    \right)    }.
\end{equation}
By introducing the Majorana fermions $ a_k= c_k +c_k^{\dagger}, b_k= i(c_k - c_k^{\dagger} )$, $U_F$ reads 
\begin{equation}
	U_F  = \prod_{k>0}  e^{   J \left[ \cos k \left( a_k b_k + a_{-k}b_{-k}     \right)  - \sin k \left( a_k a_{-k}  -b_{k} b_{-k}    \right)     \right]    }  e^{  -h \left(  a_k b_k +a_{-k} b_{-k}   \right)    }. 
\end{equation}
Defining $A_k= \begin{pmatrix} a_k& b_k & a_{-k} & b_{-k}  \end{pmatrix}^T$, one finds $ U_F  =  \prod_{k>0}  e^{   \frac{1}{4} A_k^T  W_{XX,k} A_k  }   e^{   \frac{1}{4} A_k^T  W_{Z,k} A_k  }$
where
\begin{equation}
	W_{XX,k}=\begin{pmatrix}
		0 & 2J \cos k  & -2J \sin k & 0\\
		-2J\cos k & 0 & 0  & 2J \sin k \\
		2J \sin k  &  0  &  0 &  2J \cos k \\
		0&  -2J \sin k & -2J \cos k & 0 
	\end{pmatrix}, \quad  W_{Z,k}=\begin{pmatrix}
		0 &  -2h  & 0  & 0\\
		2h & 0 & 0 &0 \\
		0  &  0  &  0 &  -2h \\
		0&  0   & 2h & 0 
	\end{pmatrix}.
\end{equation}
Since the product of two Gaussian states remains a Gaussian, $U_F$ can be simplified as
\begin{equation}
	U_F= \prod_{k>0} e^{ \frac{1}{4} A_k^TW_k A_k   },
\end{equation}
where 
\begin{equation}
	e^{W_k} =e^{W_{XX,k}}  e^{W_Z,k}.
\end{equation}
One can introduce a Floquet Hamiltonian $ -i H_{k}=  \frac{1}{4} A_k^TW_k A_k $ so that $U_F= \prod_{k>0} e^{ -i  H_k }$. Being quadratic in Majoranas, $H_k$ can be diagonalized using an orthogonal transformation on $A_k$ as $H_k=\frac{i}{2}  \varepsilon_k   \left( \gamma_k'\gamma_k'' + \gamma_{-k}'\gamma_{-k}''       \right)$, where $\varepsilon_k$ is the corresponding energy. In particular, $\varepsilon_k$ can be obtained from $w_k$ (eigenvalues of of $W_k$) through 
\begin{equation}
	\varepsilon_k= \pm  \sqrt{ - w_k^2  }.
\end{equation} 
After some algebra, one finds the eigenvalues of $e^{W_k}$:

\begin{equation}
	e^{w_k}=  \frac{x}{4}  \pm  \sqrt{ \left(  \frac{x}{4}  \right)^2-1  }
\end{equation}
where $ x=2(1+\cos k) \cos(2h-2J) + 2(1-\cos k) \cos(2h+2J)$.  Below we find that  the number of $k$ modes with purely real energy $\varepsilon_k$ is extensive in the system size $L$ when the real part of both $J$ and $h$ is $\pi /4$, which is ultimately responsible for the volume-law bipartite entanglement of long-time-evolved states. To analyze this case, we take $h=\pi/4+ i \alpha_h$ and $J=\pi/4+ i  \alpha_J$, where $\alpha_h, \alpha_J$ are real, and find 

\begin{equation}
	x= 2(1+\cos k) \cosh(2\alpha_h -2 \alpha_J)- 2(1-\cos k) \cosh(2\alpha_h +2 \alpha_J). 
\end{equation}

\begin{figure}[t]
	\begin{subfigure}[b]{0.33\textwidth}
		\includegraphics[width=\textwidth]{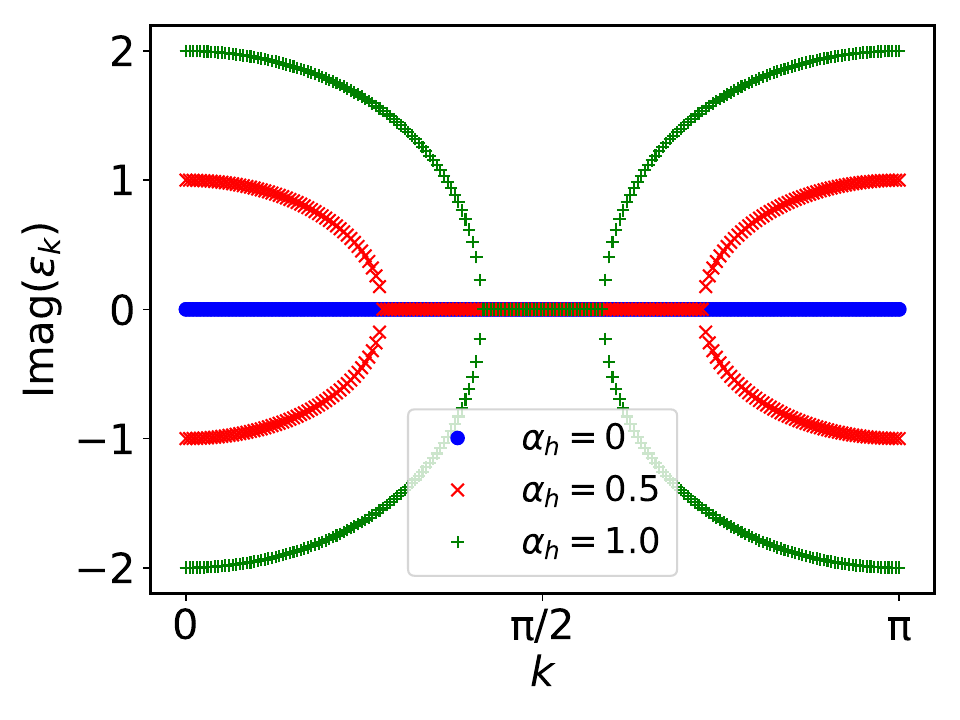}
	\end{subfigure}
	\begin{subfigure}[b]{0.33\textwidth}
		\includegraphics[width=\textwidth]{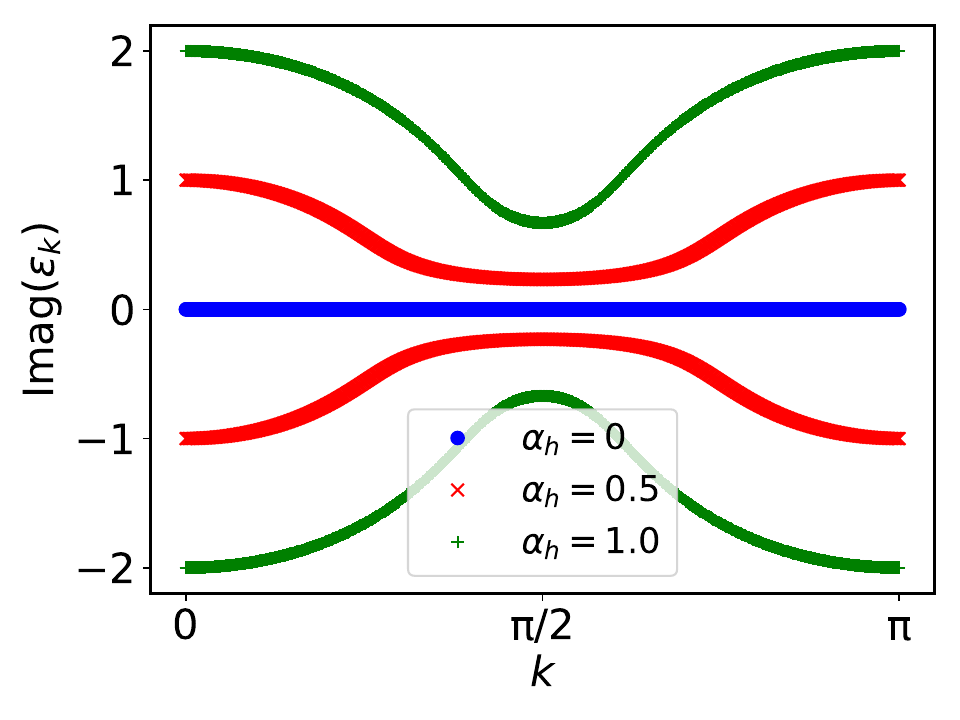}
	\end{subfigure}
	\begin{subfigure}[b]{0.33\textwidth}
		\includegraphics[width=\textwidth]{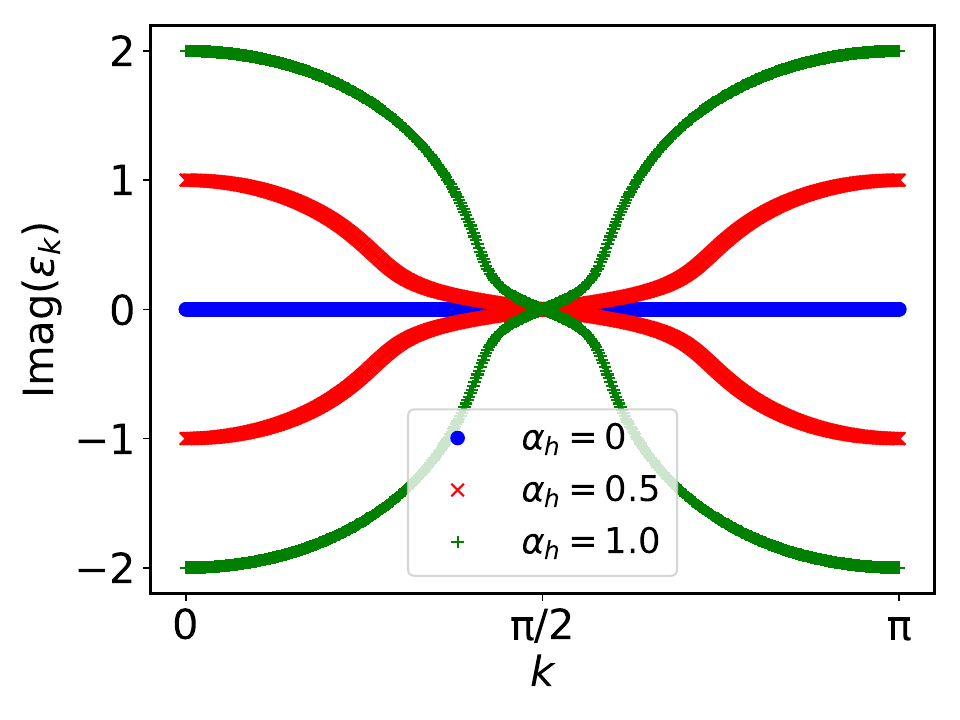}
	\end{subfigure}
	\caption{Imaginary part of the single-particle energy spectrum $\{\varepsilon_k\}$. Left: $J= \frac{\pi}{4}$,  $h= \frac{\pi}{4} +  i \alpha_h $. There exists a finite interval of $k$ modes with real energy (Eq.\ref{eq:k_modes}) for any non-infinite $\alpha_h$. Middle: $J= \frac{\pi}{4}-0.1$, $h= \frac{\pi}{4} + i \alpha_h $. Any non-zero $\alpha_h$ results in the absence of $k$-modes with purely real energy. Right: $J= \frac{\pi}{4}$, $h= \frac{\pi}{4}-0.1 +  i \alpha_h $. Only the $k$-mode with $k=\pi/2$ supports purely real energy at any non-zero $\alpha_h$.}
	\label{fig:appendix_spectrum}
\end{figure}

For $\abs{x/4}<1$, we find $e^{w_k}=\frac{x}{4} \pm i \sqrt{ 1-\left(  \frac{x}{4} \right)^2}$, and the corresponding energy $\varepsilon_k= \pm \sqrt{-w_k^2}$ is real. 

Now let's solve for the inequality $\abs{x/4}<1$ analytically for certain simple cases to identify the $k$-modes with purely real single particle energy $\varepsilon_k$. For $\alpha_J=0$, one has $\abs{ \cos k  }< \frac{1}{ \cosh(2  \alpha_h)  }$, and for any finite (i.e. non-infinite) $\alpha_h$, there is a finite interval of $k$ with purely real energy (see also Fig.\ref{fig:appendix_spectrum} left):
\begin{equation}\label{eq:k_modes}
	k  \in I_k=\left(     \frac{\pi}{2} - \sin^{-1}\left[\frac{1}{\cosh(2\alpha_h   )}  \right] ,  \frac{\pi}{2} + \sin^{-1}\left[  \frac{1}{\cosh(2\alpha_h   )}  \right]   \right). 
\end{equation}
Within the quasiparticle picture\cite{cardy_quench_2005}, since only those quasiparticle pairs with purely real energy have an infinite lifetime, Eq.\ref{eq:k_modes} implies the existence of finite density of such quasiparticle pairs, resulting in the volume-law entanglement in long-time-evolved states at any non-infinite $\alpha_h$. In particular, the volume-law coefficient of entanglement entropy follows $S_A/L_A\sim \int_{k \in I_k} dk~  s(k)$. $I_k$ (defined in Eq.\ref{eq:k_modes}) specifies the interval of $k$-modes with purely real energy. $s(k)$ is the entanglement contributed from the quasiparticle pair with momentum $k$, and is a non-universal function determined from the initial state. For large $\alpha_h$, since the length of interval $I_k$ decays exponentially as $e^{ -2 \alpha_h}$, the volume-law coefficient $S_A/L_A$ decays exponentially as well:
\begin{equation} 
	\frac{S_A}{L_A} \sim e^{ -b \alpha_h},
\end{equation}
where $b>0$ is a non-universal number that depends on the initial state.

Another simple case is $\alpha= \alpha_J =\alpha_h$, where the corresponding $k$ modes with real energy satisfy $ 0< k< k_1 =\cos^{-1} \left[ \frac{ \cosh(4\alpha)-3   }{  \cosh(4\alpha) +1 }   \right]$.  For large $\alpha$, one finds $k_1 \sim e^{-2 \alpha}$, implying the volume-law coefficient  
\begin{equation}
	\frac{S_A}{L_A} \sim e^{ - c \alpha}.
\end{equation}
where $c>0$ is a non-universal number that depends on the initial state.

Although here we only discuss two cases (varying $\alpha_h$ at fixed $\alpha_J=0$ and varying $\alpha=\alpha_h=\alpha_J$), we checked that the condition $\text{Re}(J)=\text{Re}(h)=\pi/4$ always gives extensive number of $k$-modes with purely real energy, indicating volume-law entanglement. In strong contrast, any deviation from $\text{Re}(J)=\text{Re}(h)=\pi/4$ gives $O(1)$ number of $k$-modes with purely real energy, resulting in the absence of volume-law entanglement (see Fig.\ref{fig:appendix_spectrum} middle and right).

\section{Additional data for the 2d Clifford circuit} \label{appendix:2dlogscaling}
\begin{figure}[h]
	\centering
	\begin{subfigure}[b]{0.37\textwidth}
		\includegraphics[width=\textwidth]{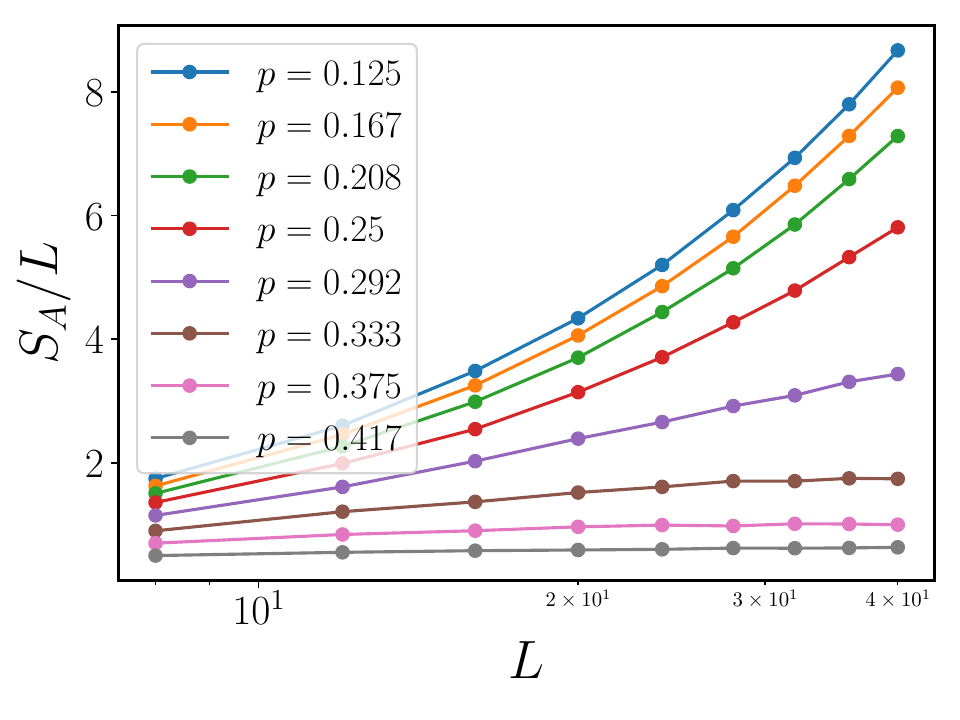}
	\end{subfigure}
	\begin{subfigure}[b]{0.37\textwidth}
		\includegraphics[width=\textwidth]{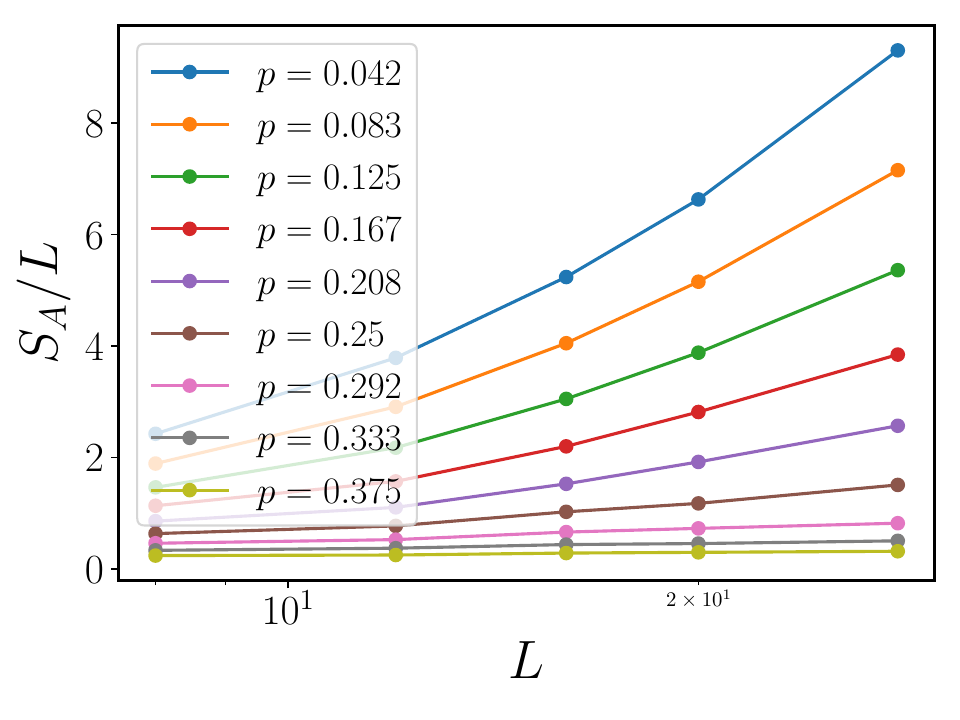}
	\end{subfigure}
	\caption{Left: Long-time entanglement entropy $S_A$ of a subregion of size $L/2\cross L$ averaged over $O(10^3)$ random realizations of the unitary circuit defined in Eq.\ref{eq:2dim}. Right: Long-time entanglement entropy of a subregion of size $L/2\cross L$ averaged over $O(10^4)$ random realizations of the rotated non-unitary circuit (Eq.\ref{eq:2dim_rotated}). The critical point is at $p_c\approx 0.28$ for both circuits.}  \label{fig:appendix_2d}
\end{figure}

Here we present additional data (Fig.\ref{fig:appendix_2d}) on the scaling of entanglement entropy for the late-time states evolved by the unitary circuit (Eq.\ref{eq:2dim}) and the non-unitary circuit (Eq.\ref{eq:2dim_rotated}).  At the critical point $p_c \approx 0.28$, the data is indicative of the scaling $S \sim L \log L$.

\section{Purification dynamics for the rotated MBL circuit defined in Eq.\ref{eq:mbl_rotated_evolve}}\label{appendix:mbl}
\begin{figure}[h]
	\centering
	\begin{subfigure}[b]{0.37\textwidth}
		\includegraphics[width=\textwidth]{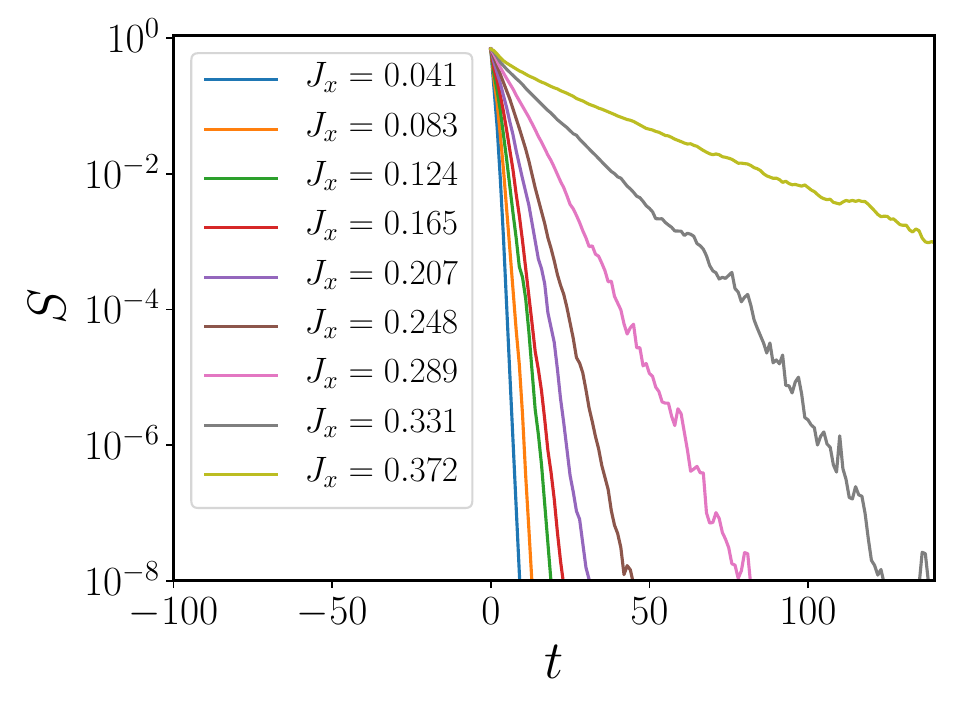}
	\end{subfigure}
	\begin{subfigure}[b]{0.37\textwidth}
		\includegraphics[width=\textwidth]{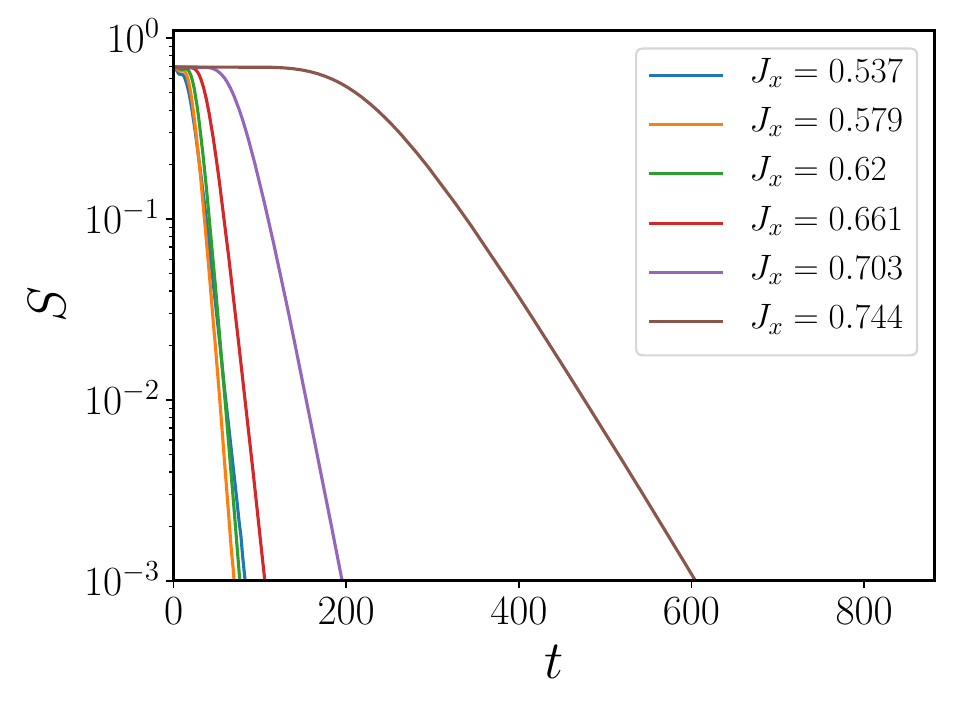}
	\end{subfigure}
	\caption{The entanglement entropy of an ancilla qubit that is initially prepared in the maximally entangled state with the system of size $L=14$, and then evolved with the non-unitary circuit in Eq.\ref{eq:mbl_rotated_evolve}. Averaging is done over 2000 realizations of the disorder.}  \label{fig:appendix_mbl_purification}
\end{figure}

Here we present additional data for the entanglement dynamics of an ancilla qubit that is initially maximally entangled with the system, and then evolved with the non-unitary circuit (Eq.\ref{eq:mbl_rotated_evolve}). For $J_x\lesssim 0.4$ (Fig.\ref{fig:appendix_mbl_purification} left), the entanglement $S$ of the ancilla qubit decays exponentially with time from its initial value, while for $J_x \gtrsim 0.4$ (Fig.\ref{fig:appendix_mbl_purification} right), $S$ remains at its initial value for  time that is superlinear in $L$ (see inset of Fig.\ref{fig:interacting_mbl}(d) for scaling with $L$), followed by an exponential decay.

 \end{document}